\documentclass[twocolumn]{aastex6}
\usepackage{graphicx}	
\usepackage{amsmath}
\usepackage{amssymb}
\usepackage{subfigure}
\usepackage{color}
\usepackage{txfonts}
\usepackage{tabto}
\usepackage{braket}
\usepackage{lineno}

\newcommand\mceq{{\sc MCEq}}

\newcommand{\be}{\begin{equation}}
\newcommand{\ee}{\end{equation}}
\newcommand{\ba}{\begin{eqnarray}}
\newcommand{\ea}{\end{eqnarray}}
\newcommand{\bc}{\begin{center}}
\newcommand{\ec}{\end{center}}

\newcommand\simlt{\lower.5ex\hbox{$\; \buildrel < \over \sim \;$}}
\newcommand\simgt{\lower.5ex\hbox{$\; \buildrel > \over \sim \;$}}

\graphicspath{{figures/}{skymaps/}}

\begin{document}

\title{Life in the Bubble: How a nearby supernova left ephemeral footprints on the cosmic-ray spectrum and  indelible imprints on life}
\author{Caitlyn Nojiri\altaffilmark{1}, No\'emie Globus\altaffilmark{1,2,3}, and Enrico Ramirez-Ruiz\altaffilmark{1}}
\altaffiltext{1}{Department of Astronomy and Astrophysics, University of California, Santa Cruz, CA 95064, USA}
\altaffiltext{2}{Kavli Institute for Particle Astrophysics and Cosmology, Stanford University, Stanford, CA 94305, USA}
\altaffiltext{3}{Astrophysical Big Bang Laboratory, RIKEN, Wako, Saitama, Japan}

\begin{abstract}
The Earth sits inside  a 300pc-wide void that was carved by a series of supernova explosions that went off tens of millions of years ago, pushing away interstellar gas and creating a bubble-like structure. The $^{60}$Fe peak deposits found in the deep-sea crust have been interpreted by the imprints left by the ejecta of supernova explosions occurring  about 2-3 and 5-6 Myr ago. It is likely that the $^{60}$Fe peak at about 2-3 Myr originated from a supernova occurring in the Upper Centaurus Lupus association in Scorpius Centaurus ($\approx$140 pc) or the Tucana Horologium association ($\approx$70 pc). Whereas, the $\approx$ 5-6 Myr peak is likely attributed to the solar system's entrance into the bubble. In this {\it Letter}, we show that the supernova source responsible for synthesizing the $^{60}$Fe peak deposits  $\approx$ 2-3 Myr ago can consistently explain the cosmic-ray spectrum and the large-scale anisotropy between 100 TeV and 100 PeV. {    The cosmic-ray knee could then potentially be attributed entirely to a single nearby “Pevatron” source}. Matching the intensity and  shape of the cosmic-ray spectrum  allows us to place stringent constraints on the cosmic-ray energy content from the supernova as well as on the cosmic-ray diffusion coefficient. Making use of such constraints we provide a robust  estimate of the temporal variation of terrestrial ionizing cosmic radiation levels and discuss their implications in the development of early life on Earth by plausibly
influencing the mutation rate and, as such,
conceivably assisting in the evolution of complex organisms.
\end{abstract}
\section{Introduction}
Life on Earth is constantly evolving under continuous exposure to ionizing radiation from both terrestrial and cosmic origin. While bedrock radioactivity slowly decreases on billion year timescales \citep{karam1999calculations,2020ApJ...903L..37N}, the levels of cosmic radiation fluctuates as our solar system travels through the Milky Way. Nearby supernova (SN) activity has the potential to raise the radiation levels at the surface of the Earth by several orders of magnitude, which is expected to have a profound impact on the evolution of life \citep{1969supe.book.....S,1995PNAS...92..235E}. In particular,  enhanced radiation levels are expected when our solar system passes near OB associations. The winds associated with these massive stellar factories are expected to initially inflate superbubbles of hot plasma, which can be the birth places of a large fraction of the core collapse explosions taking place within the OB association \citep{2018AdSpR..62.2750L}. The solar system entered such a superbubble, commonly referred to as the Local Bubble (LB), about 6 Myr ago, and currently resides near its center \citep{Zucker2022}. The presence of freshly synthesized radioisotopes detected near the Earth's surface gives credence to the idea that our Solar System has infiltrated a highly active SN region within the the Milky Way \citep{2002PhRvL..88h1101B,2023ApJ...947...58E}. Most notably, the temporal variation in the concentration of  $^{60}$Fe in sediment and crust regions  \citep{doi:10.1126/science.aax3972} places stringent constraints on the positions and ages of the closest SN events \citep{2015ApJ...800...71F,HydePecaut, Zucker2022}.

In this {\it Letter} we combine recent results on the properties  of the LB and the detection of $^{60}$Fe in deep-sea sediments to predict the cosmic-ray flux expected from a near-Earth core collapse SN event. We suggest that a single local PeVatron source, likely originating from the Scorpius Centaurus or the Tucana Horologium stellar associations,  was responsible for   producing  most of the freshly synthesized  $^{60}$Fe peak $\approx$2.5 Myr. We then proceed to calculate the associated cosmic-ray flux. This suggestion is given further credence by recent measurements of the cosmic-ray spectrum, composition, and large-scale anisotropy. Motivated by the derived source constraints, our goal is to provide a robust estimate of the temporal variations of cosmic-ray radiation doses experienced by Earth's inhabitants, using all available observational constraints.  This  {\it Letter} is organized as follows. In Section~\ref{sec:obs}, we detail recent observational results that constrain the parameters of our cosmic-ray injection model, which is described in Section~\ref{sec:mod}. Our results and their match to observational constraints are presented in Section~\ref{sec:res}. Our conclusions are submitted in Section~\ref{sec:dis}.

\begin{figure*}
\includegraphics[height=6.3cm]{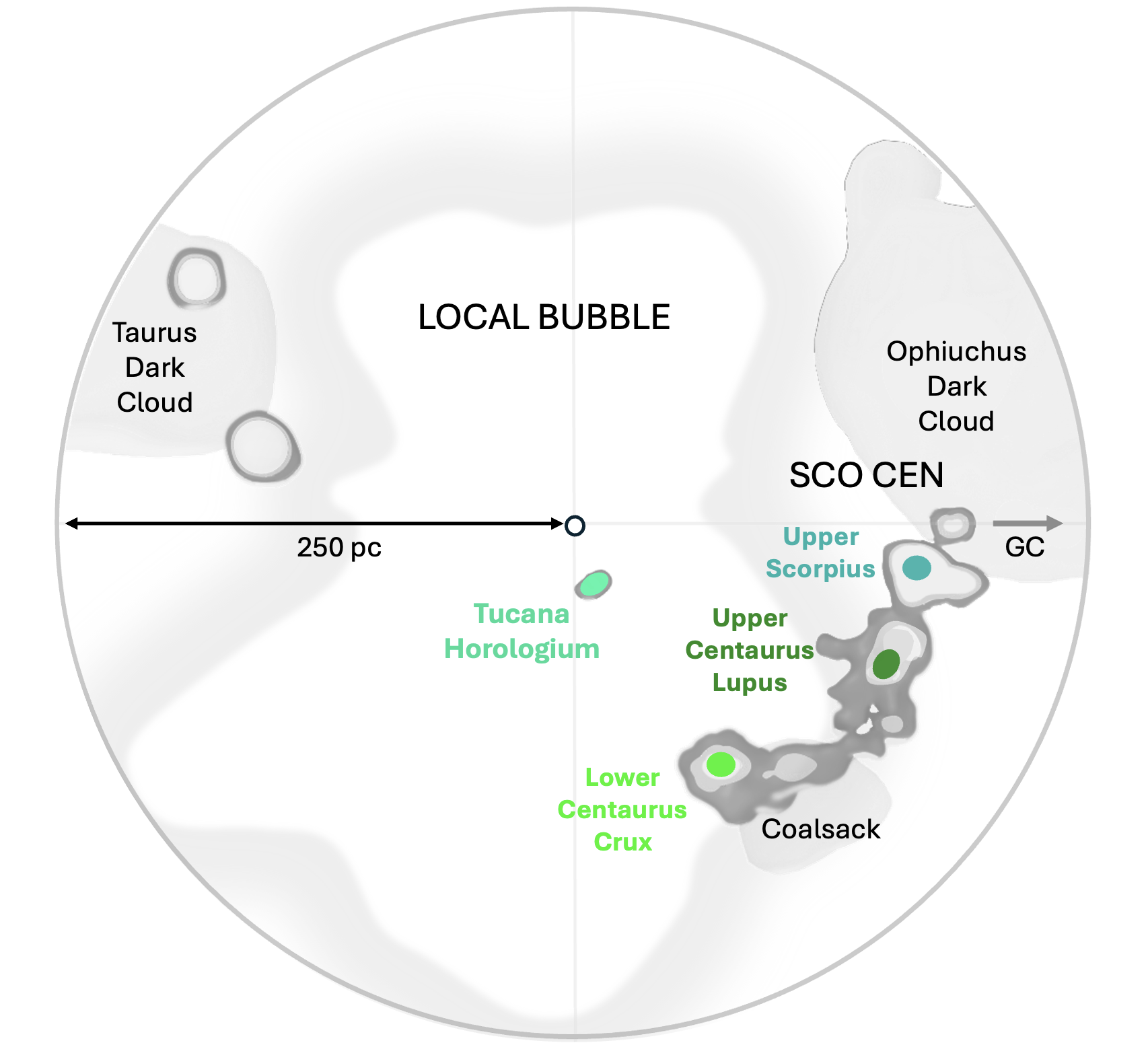}
\includegraphics[height=6.1cm]{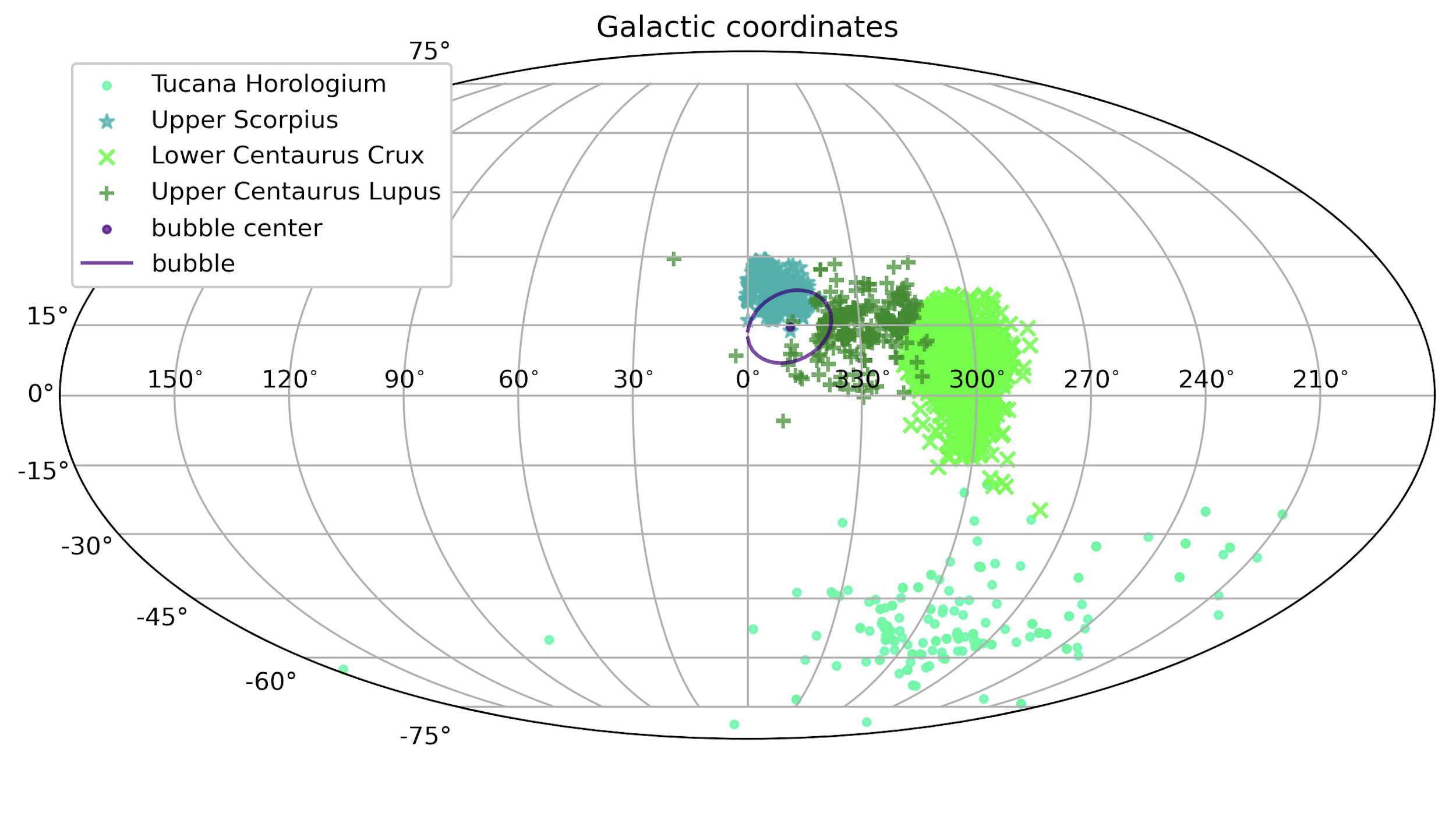}
    \caption{The configuration of stellar associations in and around the  LB. {\it Left} panel: Shown is a  {     A projection (side view, horizontal being the Galactic plane) } of today's LB and the locations of the nearby stellar associations. The shape of the LB is taken from \citep{Zucker2022,2024ApJ...973..136O}. {     The shaded region illustrates the shell of dust surrounding the LB}. The direction of the Galactic Center, GC, is denoted by an arrow.  {\it Right} panel: Shown are the positions in Galactic coordinates of the nearby stellar associations Tuc-Hor and Sco-Cen's  subgroups: Lower Centaurus Crux (LCC), Upper Centaurus Lupus (UCL), and Upper Scorpius (US). Also shown ({\it bubble}) is the new Galactic bubble discovered by \citet{2018A&A...617A.101R}, which is likely the remnant  of a supernova that took place in UCL. We anticipate an anisotropy in the distribution of arrival directions of cosmic rays that results from  SN explosions hosted by these nearby stellar associations.} 
    \label{fig:LBmap}
\end{figure*}

\section{OBSERVATIONAL CONSTRAINTS}\label{sec:obs}
In this section we present a summary of what we have learnt so far about the near-Earth massive stellar associations. We show that sufficient progress has been made in order to identify the necessary  ingredients needed to estimate the history of  cosmic-ray irradiation of our planet. The following questions were used to guide the  assembly of these essential ingredients into a general model scheme.

\subsection{What nearby stellar associations are thought to dominate the ongoing SN activity?}      
The nearest active star-forming region to the Sun is the $\approx$15-Myr-old OB association Scorpius Centaurus (Sco-Cen). It is thought to  be responsible for most of the massive stellar  activity that conceived the LB \citep{Frisch2011,Zucker2022}. The associated SN activity in Sco-Cen is also credited with creating the Loop I superbubble which has been observed to interact and subsequently merge with the LB \citep{1995A&A...294L..25E}. Not to mention that our solar system is conjectured to be currently 
traversing an outflow originating from Sco-Cen \citep{2024arXiv240713226P}.

At present OB Sco-Cen association spans distances between 100 and 150 pc and includes several molecular clouds currently undergoing star formation \citep{2023A&A...678A..71R}.   
Sco-Cen is divided into three  subgroups: the Lower Centaurus Crux (LCC), Upper Centaurus Lupus (UCL) and Upper Scorpius (US). Another highly viable candidate for hosting a near-Earth SN is the $\approx$40~Myr-old Tucana-Horologium \citep[Tuc-Hor,][]{2023MNRAS.520.6245G,HydePecaut} association,  one of the closest young stellar groups to the Solar System with an average distance of about 46~pc. A  layout ({\it left} panel) and a sky map  ({\it right} panel) of the LB  and nearby stellar associations is shown in Figure~\ref{fig:LBmap}.  We foresee an anisotropy in the distribution of arrival directions of cosmic rays associated  with a SN explosion residing in these nearby stellar associations.  

\subsection{Which  stellar association is most likely to be responsible the latest  SN event?} 

The LB is expected to have been inflated by a combination of stellar winds and SN explosions \citep[e.g.,][]{2014MNRAS.442.2701R}. These SN blast waves can disperse freshly synthesised elements that can then be turbulently mixed \citep{2020ApJ...899L..30G,2022ApJ...936L..26K,2023ApJ...949..100K} throughout the LB. This is expected  to be the case for the freshly synthesised dust composed of proto-silicates, silicon dioxide and iron oxide, containing the radioactive isotope $^{60}$Fe that was captured  by the Earth and incorporated into the geological record. Peak concentrations of $^{60}$Fe occurred about 2-3 Myr ago and 6-7 Myr ago \citep{doi:10.1126/science.aax3972}. 

Constraints on the birth-site of the supernova progenitors responsible for these two main peaks can be placed from the initial mass function (IMF), the ages of nearby stellar groups and the metal dispersion time across the LB. The transport  timescales expected if  $^{60}$Fe was entrained in the supernova blast wave plasma are $\lesssim$ 0.1 Myr \citep{2022ApJ...936L..26K} and $\lesssim$ 1 Myr  if  $^{60}$Fe arrived in the form of supernova dust, whose dynamics differ from but are connected to the evolution of the blast wave material \citep{2023ApJ...947...58E}. As such, it is believed that the radioactive age can be effectively used to constraint the  time since explosion.    Sco-Cen  entered the LB $\approx$10 Myr ago \citep{2001ApJ...560L..83M, 2006MNRAS.373..993F,2023A&A...678A..71R} and since then, a handful of SN explosions have been predicted to occur in this association \citep{2006MNRAS.373..993F}. According to \citeauthor{HydePecaut}, UCL and LCC remain plausible sites for hosting the event responsible for producing the $^{60}$Fe peak concentration 2-3 Myr ago. Both LCC and UCL contain  prospective progenitors  with  initial mass  estimates $\gtrsim 20$$M\textsubscript{\(\odot\)}$.  

Another hint for a progenitor site  relates to the discovery of a new Galactic “bubble", of radial extend 45 pc located  at a distance of $\approx$140 pc in the UCL \citep{2018A&A...617A.101R}. This remnant is  shown in Figure~\ref{fig:LBmap}, and can be identified under the label “bubble". The radial extension of the remnant  is consistent with one SN going off inside the rarefied LB medium \citep{1977ApJ...218..377W}. Not only that, the $\approx$3~Myr old runaway pulsar PSR J1932+1059 and the  runaway O star $\zeta$~Oph are both likely associated with a SN event in the UCL. Both stellar objects are estimated to have  left the UCL subgroup about 3 Myr ago \citep{2000ApJ...544L.133H}, {    although revised estimates by \citet{2020MNRAS.498..899N} indicate that could have been released $\approx 2$ Myr ago (albeit their analysis does not take into account the motion of all stellar associations within the LB).} All these observations give credibility to the idea that SN activity  in the UCL could have been responsible for producing the $^{60}$Fe peak 2-3 Myr ago. 

The stellar cluster Tuc-Hor, on the other hand,  is expected to have produced about a  single SN since the Sun entered the LB  about 6 Myr ago. A Salpeter IMF  predicts $\approx$1 star with mass $>8$~M$_\odot$ which would have evolved into core collapse  in the recent past \citep{2016IAUS..314...21M}. Tuc-Hor association can not be ruled out as a candidate of the 2-3 Myr or the 6-7 Myr $^{60}$ Fe peaks.  However, since Tuc-Hor is the oldest, \citeauthor{HydePecaut} suggest the UCL association as the most likely site for the 2-3 Myr  $^{60}$Fe peak. Another possible explanation for the 6-7 Myr $^{60}$Fe peak is attributed to the Solar System  traversing the denser shell region of the LB \citep{Fang_2020}. In absence of more stringent constraints, in what follows we consider both the Tuc-Hor cluster and the Sco-Cen's UCL subgroup as likely candidates for the production of the 2-3 Myr $^{60}$Fe deposits.

\subsection{What can be learned from cosmic-ray data?} 
One of the key  features in the cosmic-ray spectrum is the “knee” observed at around $\approx$5~PeV, where the power spectral index changes from $2.7$ to $3.3$. The presence of a clear succession of “heavy knees"\footnote{A proton knee observed  at $\approx$5~PeV, followed by a Helium at 10 PeV \citep{2024PhRvD.109l1101A}, a Silicon-like peak at around 50 PeV, and an Iron at around 100 PeV  \citep{2011PhRvL.107q1104A, 2024JCAP...05..125K}. The latest knee at $\approx$400~PeV marks the end of the ultra-heavy cosmic-ray component \citep{2003APh....19..193H}.} at high energies suggest that a source with a single maximal rigidity\footnote{Rigidity in good approximation for ultra-relativistic particles can be defined as $R = \frac{E}{Z}$, where $E$ is the total energy and $Z$ is the charge.} (5-6 PV) is dominating the spectrum in this region, just before the transition to the extragalactic cosmic ray contribution taking place at around 100 PeV \citep{2015PhRvD..92b1302G}. This is supported  by the phase-flip in the dipole anisotropy at around 100 TeV in the Galactic center direction \citep[e.g.,][]{2024arXiv240108952F}. Motivated by this, we  surmise that a lone PeVatron source should be able to explain the spectrum, composition, and anisotropy in the 100 TeV-100 PeV range, and conjecture that the same source was likely the same  SN responsible  for producing the 2-3 Myr $^{60}$Fe peak. In fact, PeV acceleration is alleged to be efficient during the early SN stages \citep[e.g., ][]{2021Univ....7..324C} and   SN remnants exploding in hot bubble environments have been proposed as viable candidate sources for energies above ~PeV  \citep{Vieu2023}.

Our LB is not unique. The interstellar medium in the Milky Way disk is filled with plenty of superbubbles, believed to be the remnants of past collective supernova activity. The studies of these superbubbles can help shed light on how our LB was carved.  Recently, {\tt LHAASO} discovered a giant $\gamma$-ray bubble structure in the Cygnus star-forming region with photon energies above 100 TeV, clearly indicative of  acceleration of protons up to PeV energies in a region containing a massive stellar  OB association \citep{2024SciBu..69..449L}.  The total energy cosmic-ray content  in all the cosmic rays presently filling the Cygnus superbubble is constrained from  {\tt Fermi} observations to be  1.3-6.5 $\times 10^{49}$ erg \citep{2011Sci...334.1103A}. Individual SN remnants are consistent with similar  cosmic-ray energy content based on $\gamma$-ray observations \citep{1999ICRC....3..480A}. On that account, in what follows, we consider a total cosmic-ray injection energy per SN  $\geq 10^{49}$ erg. 

\subsection{How far from Earth were the clusters progenitor candidates at the time of the SN explosion?} 
\begin{figure}[!h]
\includegraphics[height=6.7cm]{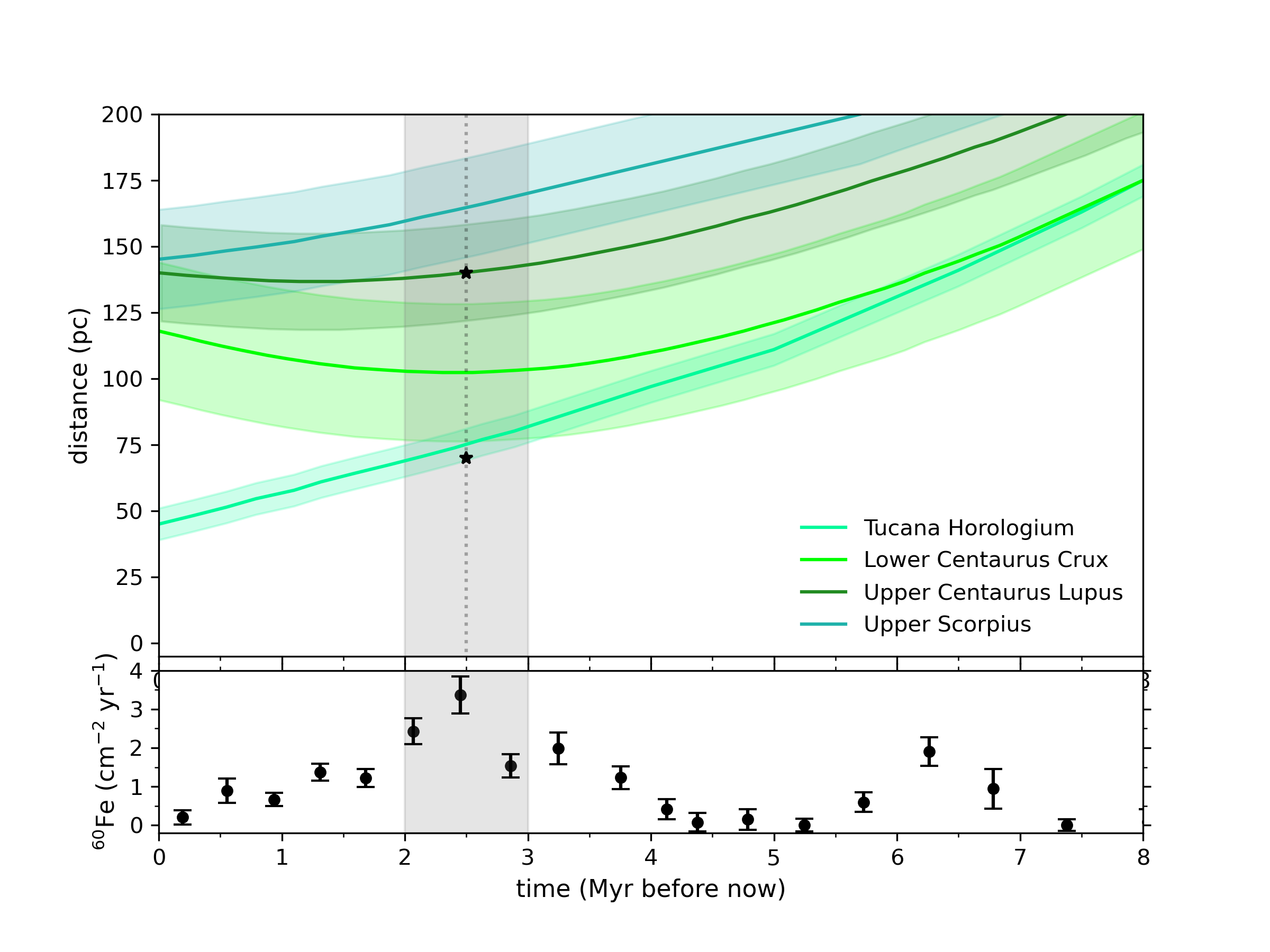}
    \caption{The distance from Earth to the clusters progenitor candidates. {\it Upper panel:}  The evolution of our distance to the various stellar associations in time, which has been adapted from \citet{2002PhRvL..88h1101B}. The distance spread of the stellar associations is shown as shaded regions, which  correspond to the associated $2\sigma$ uncertainties in the distance estimate. The color code for the different associations is the same as in Figure~\ref{fig:LBmap}. {\it Lower panel:} The collected data of $^{60}$Fe, which shows the two distinct peaks and has been adapted  from  \cite{doi:10.1126/science.aax3972}. The shaded vertical grey region shows  the most recent peak concentration at about 2-3 Myr. The width in peak times is thought to be produced by the transport timescales expected for  $^{60}$Fe supernova dust \citep{2023ApJ...947...58E}.  The two black star symbols  show the two SN progenitor candidates we consider in light of the observational constraints. These are located at a distance from Earth $r_{\rm inj}=$70~pc at $t_{\rm inj}=$2.5~Myr in Tucana Horologium , and $r_{\rm inj}=$140~pc at $t_{\rm inj}=$2.5~Myr in Upper Centaurus Lupus.}\label{fig:distances}
\end{figure}

Over the past $\approx$6 Myr, the sun traveled through the LB and, as such,  the distance of the embedded stellar clusters to Earth has evolved with time. In order to understand the evolution of the clusters in relationship to the Earth we use data from \citet{2002PhRvL..88h1101B} and \citet{2023A&A...680A..39S}. The distance of the centers of the stellar associations as a function of time are shown in Figure~\ref{fig:distances}. The uncertainties in the distance range estimates  for LCC, US, UCL, and Tuc-Hor were used to calculate the corresponding uncertainties assuming Gaussian distributions, with the shaded regions in Figure~\ref{fig:distances} corresponding to $\pm 2\sigma$. At 2.5 Myr, the core of Tuc-Hor was $\approx$70 pc away while the core  of UCL was $\approx$140 pc.  In what follows, these are the distance estimates  we consider for the SN event that produced the early $^{60}$Fe deposits.

\section{Model Assumptions and Methods}\label{sec:mod}
\subsection{Assumptions}
Motivated by the recent mapping of the star forming regions within the LB, we consider  either Tuc-Hor or  Sco-Cen's UCL subgroup as likely hosts for the 2-3~Myr $^{60}$Fe SN event. We consider that this event acted as a PeVatron  source, and is also responsible for the “knee" feature in the cosmic ray spectrum. We then make  the  natural assumption that all accelerated nuclei have the same spectrum in rigidity and simply determine the relative abundances from the observational data.  The light (p+He) data is taken from DAMPE \citep{2024PhRvD.109l1101A}, EAS-TOP \citep{2004APh....20..641A}, ARGO-YBJ  \citep{2015PhRvD..92i2005B}, and KASCADE \citep{2005APh....24....1A}. 
The heavy (Si+Fe) data is taken from  \citet{2023arXiv231205054K}. We  use data from HAWC \citep{2022arXiv220814245M}, Tibet-III \citep{2008ApJ...678.1165A} and KASCADE-Grande \citep{2023arXiv231205054K} to constraint the all-particle spectrum. The source spectrum needs to be steep enough so that the  contribution of the local PeVatron does not exceed $\lesssim$10\% of the flux at 1~GeV (Moskalenko, private com.) 

To model the cosmic-ray transport from the SN to Earth, we need to understand the properties of the magnetic field in the LB. Many attempts have been made to constrain the local diffusion coefficient from secondary cosmic ray data. Below 200 GV rigidity, the data can be well-accounted for with a single power-law form of the diffusion coefficient:
$D(R)=D_0 (R/10\rm GV)^\eta$, where the parameter $\eta$ governs the evolution with rigidity, which is related to the underlying assumption on the turbulence spectrum. 

{    Recent observations of the so-called {\it TeV halos}  has hinted that the diffusion coefficient in tens of parsecs around nearby pulsars is inferred to be two order of magnitude lower when compared to the interstellar medium \citep{2017Sci...358..911A}. These authors find $D(100~{\rm TeV}) = 4.5\times10^{27}$ cm$^2$~s$^{-1}$, implying $\approx$$10^{26}$  cm$^2$~s$^{-1}$ at 10 GV for a Kraichnan spectrum. \citet{2012ApJ...761...17H} estimated the diffusion coefficient in the outer heliosheath to be consistent with several $10^{26}$ to $10^{27}$ cm$^2$~s$^{-1}$ at 1~GV which suggests it can be as low as $10^{27}$ cm$^2$~s$^{-1}$ at 10~GV. This reveals that the propagation of cosmic rays can be altered by their self-generated turbulence, as pointed out by several authors \citep[e.g.,][]{2019MNRAS.484.2684N,2022MNRAS.512..233S}. Lastly, the parameter $\eta$  has been recently constrained by the DAMPE Collaboration to be $\approx$ 0.477, which is very close to the prediction of the Kraichnan theory of turbulence \citep{2022SciBu..67.2162D}.}

Motivated by {    the above} findings, our model assumptions are listed below.
\begin{itemize}
\item  We consider two progenitor candidates: a SN with  explosion coordinates ($t_{\rm inj},r_{\rm inj}$) = (2.5~Myr, 70~pc) for  Tuc-Hor, and a SN with explosion coordinates (2.5~Myr, 140~pc) for UCL. We assume that the explosions inject $\approx5\times 10^{49}$~erg in cosmic rays.\footnote{    We allow variations by a factor $\approx$50 around this value to fit the knee, depending on our specific assumption for the choice of the diffusion coefficient at 10 GV.} 
\item We assume all nuclei have the same spectrum in rigidity $dN/dR\propto R^{-\alpha}$  up to an exponential cutoff in ${\rm e}^{-R/R_{\rm cut}}$ with $R_{\rm cut}\approx 5$~PV to fit the knee. From this, we directly determine the slope, $\alpha$, and the relative abundances of the different elements from the observational data.
\item  We assume a functional form for the diffusion coefficient $D(R)=D_0\,(R/10\rm GV)^\eta$, {    with values of $D_0$ between $10^{27}$ and $10^{29}$ cm$^2$ s$^{-1}$ and $\eta=0.5$. }
\end{itemize}
In the sections that follow we describe in detail the components of the  numerical model used  to calculate the cosmic-ray transport from the source to the Earth and then the radiation doses experienced at various atmospheric depths. To do this, we first calculate  the flux at the top of the atmosphere by solving the diffusion of the primary cosmic rays.   

\subsection{Cosmic-ray intensity at the top of the atmosphere}
The cosmic ray intensity contribution is given by 
\be
   J_p(E,r,t)= \frac {c}  {4\pi} \;  f,
\ee
where $f(E,r,t)$ is the distribution function of protons at
time $t$ and radial distance $r$ from the source, which, in turn, 
satisfies the radial-temporal-energy dependent diffusion equation 
\ba
   && ({\partial f}/{\partial t})=({D(E)}/{r^2}) ({\partial}/{\partial
   r}) r^2 ({\partial f}/{\partial r}) +\nonumber \\  
   && \hspace{4cm} ({\partial}/{\partial
   E}) \, (Pf)+Q,
   \label{DL2}
\ea
where we use spherical coordinates, with $r$ being  the radial
distance from a given accelerator, $P$ representing the energy losses and $Q$ the injection term for cosmic rays. We assume that the proton energy loss $P$ is due to nuclear interactions.  The nuclear loss rate is $P_{\rm nuc} = E/\tau_{\rm pp}$, with $\tau_{\rm pp}=(n_{\rm p} c \kappa \sigma_{\rm pp}) ^{-1}$ is the timescale for the corresponding nuclear loss, $\kappa \approx 0.45$ is the inelasticity of the interaction and $\sigma_{\rm pp}$ is the  {    total} cross section for $pp$ interactions.  Above $E_{\rm lab} = 3$ GeV, $\sigma_{pp}$ can be written  as $\sigma_{\rm pp}(E_{\rm lab})=30.364-1.716\log(E_{\rm lab}) +0.981 \log(E_{\rm lab})^2\,\,{\rm mb}$, assuming EPOS-LHC for the hadronic interaction model. 

A solution to the diffusion equation for an arbitrary energy loss term, a fixed diffusion coefficient, and an impulsive  injection spectrum $f_{\rm inj}(E)$,  such that  $Q(E,r,t) = N_0 f_{\rm inj}(E) \delta(r_{\rm inj}) \delta(t_{\rm inj})$, can be found for the particular case in which $D(E)\propto E^\eta$ and $f_{\rm inj}\propto E^{-\alpha}$. Under such conditions, the solution to the diffusion equation \citep{1996A&A...309..917A} can be written as
\ba
   f(E, r,t) \approx \frac{N_0 E^{-\alpha}}{({\pi^{3/2} r_{\rm dif}^3}) }
   \exp \left[  - {(\alpha-1)t}/{\tau_{pp} }- ({r}/{r_{\rm diff}})^2 \right],
  \label{sol}
\ea
 where 
\be 
r_{\rm diff} = 2 \sqrt{ D(E) t \frac{\exp(t \eta / \tau_{pp})-1}{(t \eta / \tau_{pp})}}
\label{rdii}
\ee
denotes  the radius of the sphere up to which the particles of energy $E$ have time to diffuse after being  injected. {    We note that in the absence of pion production, this reduces to the well known formula $f(E, r,t) = \frac{N_0 E^{-\alpha}}{({\pi^{3/2} r_{\rm dif}^3}) }{\rm e}^{- ({r}/{r_{\rm diff}})^2}$ with $r_{\rm diff} = 2 \sqrt{ D(E) t}$. This is the case because the density of the LB is $\ll$ 0.1 proton cm$^{-3}$, which implies that cosmic ray destruction time is much longer when compared to the diffusion (or escape) time. In order to consider a high rigidity cutoff of 5 PV, we multiply the source rigidity spectrum of the different elemental contributions by ${\rm e}^{-R/R_{\rm cut}}$.} We use this solution to calculate the cosmic ray spectrum at the top of the atmosphere for our two progenitor  candidates at various times following the SN explosion. We then use these spectra as an input to calculate the radiation doses experienced at different depths in the atmosphere and underground. We describe the dose calculation procedure in the following section.

 \subsection{Atmospheric and underground fluxes}
Here we follow the same procedure as the one described in \citet{2021ApJ...910...85G}. The differential particle fluxes $\Phi(E,X)$, where $E$ is the kinetic energy and $X(h) = \int_h^\infty{\rm d}\ell~\rho(\ell)$ is the slant depth in g/cm$^2$, are computed using the one-dimensional cascade equation solver \mceq{}\footnote{\url{https://github.com/afedynitch/MCEq} (Version 1.3)} \citep{2015EPJWC..9908001F} with a kinetic energy range down to  10 MeV. The numerical routines from \citet{2019ICRC...36..961M} are used to calculate the electromagnetic cross sections. Ionization losses $\langle {\rm d}E/{\rm d}X \rangle$ are based on tables from \citep{2020PTEP.2020h3C01P,ionization_tables} and are tracked for each charged particle. 

The present flux of cosmic rays is considered to be isotropic and represented by the Global Spline Fit (GSF) \citep{2017ICRC...35..533D}, which is a modern parameterization of the cosmic ray flux between rigidity of a few GV and the highest observed energies at Earth.  The omission of the planetary magnetic field and solar modulation effects affecting the spectrum below a few GV/nucleon is not expected to qualitatively change our results.

\begin{figure*}[!h]
\includegraphics[height=16.5cm]{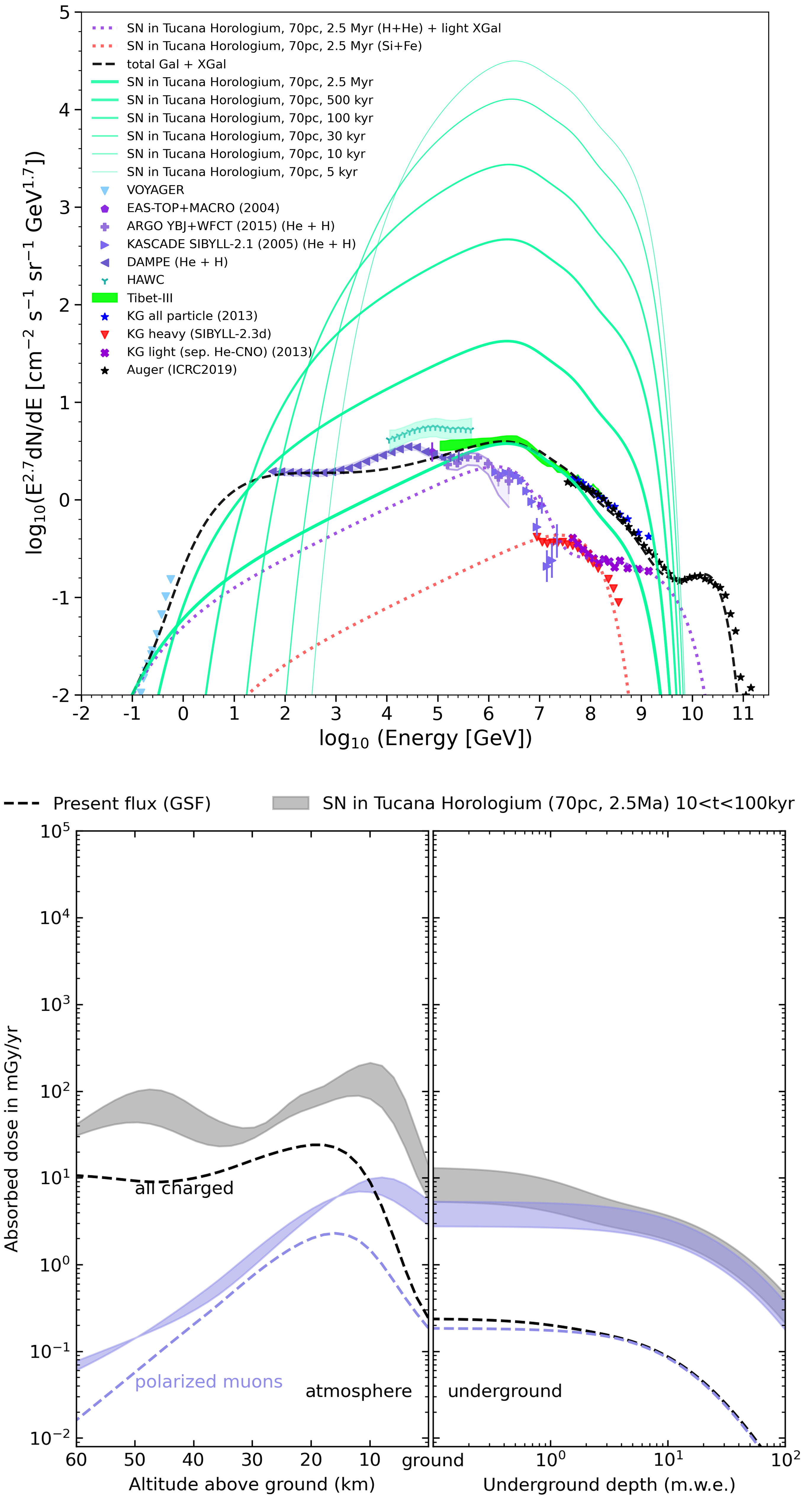}
\includegraphics[height=16.5cm]{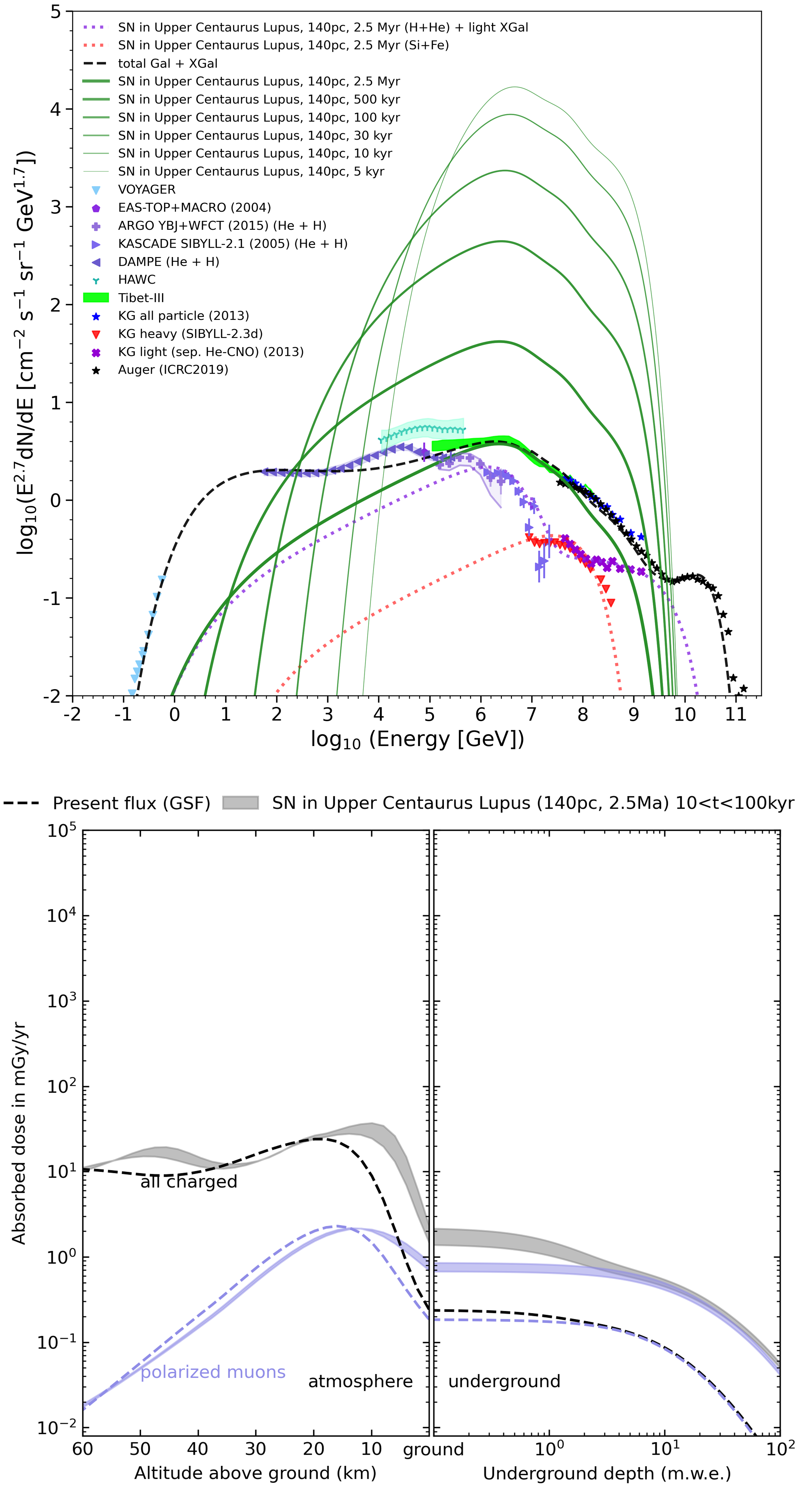}
    \caption{The cosmic-ray spectrum ({\it upper panels}) and the corresponding radiation dose experienced at different altitudes/depths ({\it lower panels}) from a supernova explosion associated with the 2.5Myr  old $^{60}$Fe peak deposits. The figures show our results for  the two different associations (same color code for the associations as in Figure~\ref{fig:LBmap}).  {\it Left panels}: PeVatron in Tucana Horologium (2.5 Myr, 70pc). {\it Right panels}: PeVatron in Upper Centaurus Lupus (2.5 Myr, 140 pc). The diffusion coefficient assumed here is $10^{27}$ cm$^2$ s$^{-1}$ at 10 GeV and follows a Kraichan energy dependence ($\eta=0.5$). Our best fit to the observed spectra is given by the  {\it dashed} and  {\it dotted} lines in the  upper panels for the two distinct SN sites ({\it left} and {\it right} panels). The reader is refer to the text for a discussion of all the separate contributions expected to the total cosmic-ray spectra ({\it dashed black} line). {    The parameters for the source spectrum are: $\alpha=1.7$, $R_{\rm cut}=$ 5~PV, $N_0=10^{49}$~erg for the SN at 70 pc  and $N_0= 4\times10^{49}$~erg for the SN at 140pc. A composition of $\approx$ 90\% light (H+He), $\approx$ 8.8\% CNO, $\approx1$\% Si+Fe, and $\approx 0.2$\% r-process elements, provide a good fit to the elemental  contributions in the 100 TeV-100 PeV energy range. We get similar results when considering an explosion at 3 Myr with $N_0=1.3\times10^{49}$~erg for the SN at 70 pc  and $N_0=5\times10^{49}$~erg for the SN at 140 pc.}}
    \label{fig:doses1}
\end{figure*}

The absorbed dose rate $d$ in Gy/s is calculated from the differential fluxes $\Phi$ with the default units (GeV cm$^2$ s sr)$^{-1}$ using
\begin{equation}
\label{eq:dose_equation}
\nonumber
    d(X) = 2\pi \sum_{p}\int_{\cos{\theta_{\rm max}}}^1 {\rm d}\cos{\theta}\int {\rm d} E_p~\Phi_p(E_p,X) \left \langle \frac{{\rm d}E_p}{{\rm d}X} \right \rangle (E_p),
\end{equation}
where $\theta_{\rm max} = \max(\pi/2,~\pi - \arcsin[r_\earth/(r_\earth + h)])$. Here  $h$ is the altitude above ground and $r_\earth$ is the radius of the planet. The index $p$ iterates over the various particle species, while the expression for $\cos{\theta_{\rm max}}$ takes into account that at higher altitudes, particle cascades can develop upwards relative to the horizon. The model of the Earth's atmosphere used here  is the \citet{us_std_atmosphere}. The reader is refer to \citet{2021ApJ...910...85G} for further details.

\section{RESULTS}\label{sec:res}

\begin{figure*}[!h]
\includegraphics[height=10.5cm]{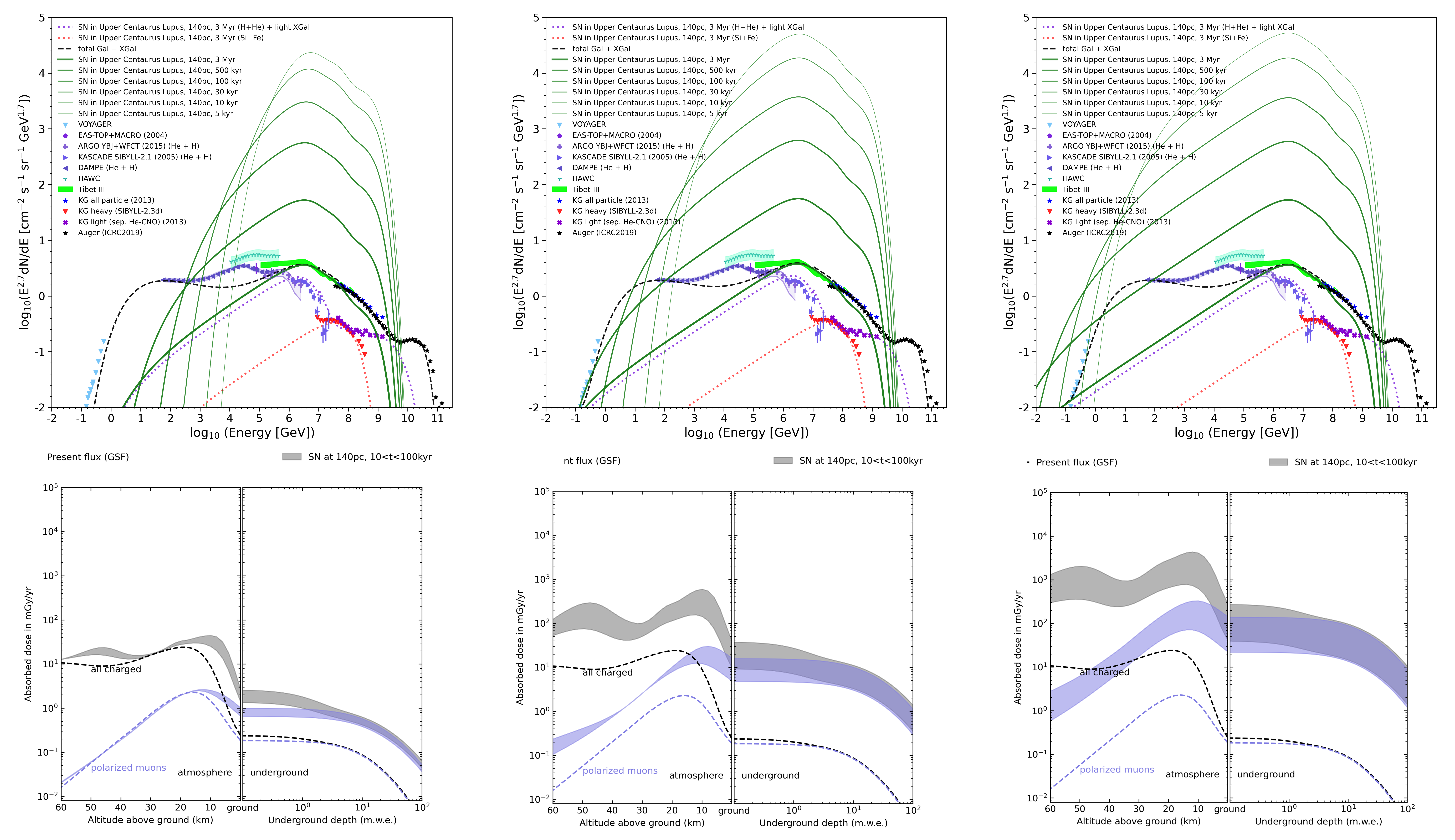}
    \caption{      Same as Fig.~\ref{fig:doses1}  for a PeVatron in Upper Centaurus Lupus (3.0 Myr, 140 pc) but varying the value of $D_0$ at  10 GeV. A good description of the observations can be achieved  for a spectral index $\alpha=1.6$ and a composition $\approx$ 91.25\% light (H+He), $\approx$ 8\% CNO, $\approx0.6$\% Si+Fe, and $\approx 0.15$\% r-process elements. {\it Left panels}: $D_0=10^{27}$ cm$^2$ s$^{-1}$ ($N_0\approx 3\times10^{48}$~erg), {\it Middle panels}: $D_0=10^{28}$ cm$^2$ s$^{-1}$ ($N_0\approx1\times10^{50}$~erg), {\it Right panels}: $D_0=10^{29}$ cm$^2$ s$^{-1}$ ($N_0\approx3\times10^{51}$~erg).}
    \label{fig:doses2}
\end{figure*}

Figures~\ref{fig:doses1} and \ref{fig:doses2} present the cosmic-ray spectra ({\it upper panels}) and the corresponding radiation doses ({\it lower panels}) experienced at different altitudes/depths on Earth. {    The cosmic-ray spectra are shown as a function of time, from  explosion  to  present day. The parameters used for the calculation are indicated in the caption. The spectra at all times have the same high rigidity cutoff for the elemental contribution. A feature common to all the spectra is the presence of a low rigidity cutoff at rigidity $R_{\rm low cut}= [r_{\rm source}/(4 t D_0)]^{1/\delta}$. The low-rigidity cutoff is due to the fact that at a given moment, only particles of sufficiently large rigidity had enough time to propagate from the source to Earth. The corresponding cutoff  moves to lower energies as time evolves, because particles with lower rigidities have time to effectively propagate.}

It can be seen that the light and heavy components of our  model provide a reasonable  description of the data in the 100 TeV-100 PeV energy range. At lower and higher energies, other sources, as expected, dominate the flux as suggested by the anisotropy data \citep{2024arXiv240108952F}. We do not aim to account for the proton “10 TeV bump" which has been recently suggested to be a re-acceleration feature by local stellar winds \citep{2022ApJ...933...78M}.  As can be seen from Figures~\ref{fig:doses1} and \ref{fig:doses2}, our PeVatron starts to dominate the flux  $\gtrsim 100$~TeV and its contribution ends at the ankle. Values of $\alpha$ around $\approx$1.6-1.7 allow the PeVatron component to become very low to negligible at 10 TeV.

The  “light ankle” observed by KASCADE-Grande can be naturally understood as the emergence of a light extragalactic component, taking over at around 100 PeV \citep{2015PhRvD..92b1302G}. The {\it dotted purple} line  is used to show the combined light PeVatron+extragalactic component, which nicely reproduces the light ankle around 100 PeV.  The {\it dotted red} line show  the fit to the heavy knee which constrain the Si+Fe contribution of our local PeVatron source to be $\lesssim$~1\%.  In both spectral figures, the {\it black dashed} line depicts the entire combined cosmic ray energy spectrum. 

 Lastly, a straightforward fit to the lower and higher (extragalactic) energy components  of the cosmic-ray spectrum, which we add to our PeVatron contribution, allows us to estimate  the total average cosmic radiation dose experienced at Earth ({\it bottom} panels in Figures~\ref{fig:doses1} and \ref{fig:doses2}). The absorbed dose rates are given in units of mGy/yr ($1 {\rm mGy/yr} \approx 3.2 \times 10^{-11}$ watt/kg).
The upper edge of the {\it grey shaded area} in the lower panels of Figures~\ref{fig:doses1} and \ref{fig:doses2} represents the average dose rate during the first $10^4$ years after explosion and the lower edge during the first $10^5$ years. We also show in {\it violet} colors the polarized muon component which is the dominant contribution of the cosmic radiation dose at ground level and is 100\% of the dose after 10 meter-water-equivalent (m.w.e.) below the surface.  
In the section that follows we discussed the consequences that this most recent SN event would have had in terms of past climate and, in particular, in the evolution of organisms on Earth.

\begin{figure}[!h]
\includegraphics[height=10.3cm]{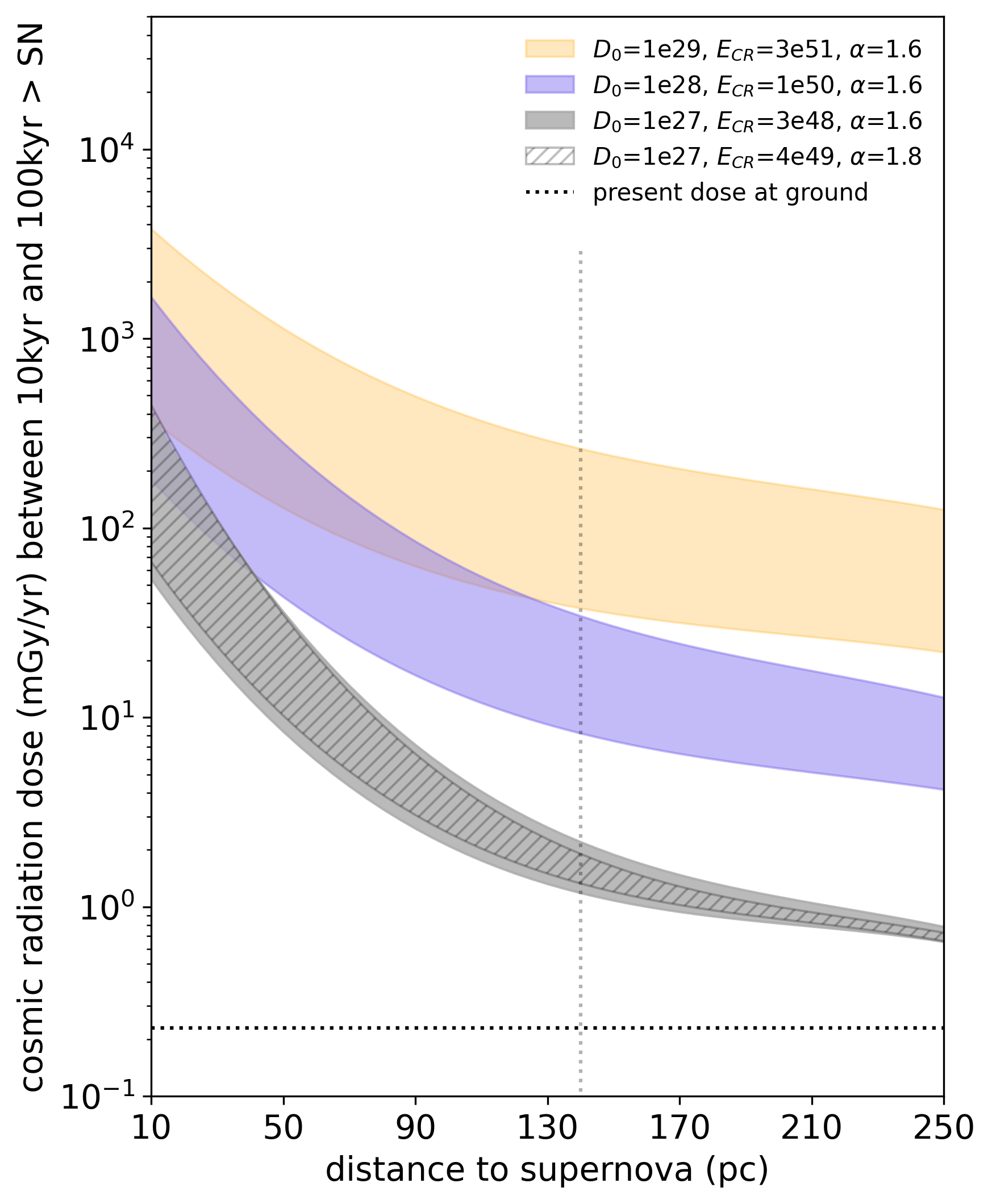}
    \caption{{    The average dose rate experienced at ground level as a function of the distance to the nearby PeVatron.  The value of the diffusion coefficient at 10 GV, the total energy released in cosmic rays and the spectral index giving a good fit to the spectrum  for a PeVatron (3 Myr, 140 pc) are indicated by the legend (see corresponding spectra in Fig.~\ref{fig:doses2}). Assuming the same parameters but just changing the distance, we calculated the doses at different distances. The shaded region represents the  average dose  between  the first 10 kyr and  the first 100 kyr.}} 
    \label{fig:dosedist}
\end{figure}

\section{DISCUSSION}\label{sec:dis}
The geological record regarding  the variations in $^{60}$Fe concentration indicate that a SN exploded near the Earth about 2.5 Myr ago. The discovery of a new “bubble" remnant  and a runaway star give credence to the idea that this SN explosion likely originated from the Upper Centaurus Lupus, a subgroup of the OB association Sco-Cen, 140 pc away in the direction of the Galactic center. 

Interestingly, the phase-flip anisotropy at 100 TeV in the direction of Upper Centaurus Lupus (Figure~\ref{fig:LBmap}) indicates, as we argued here, that the SN  that synthesized the  $^{60}$Fe, was likely also a  PeVatron source. A lone PeVatron source in the OB association Sco-Cen  would, at the same time, dominate the flux in the 100 TeV - 5 PeV range today and be responsible for synthesizing the  $^{60}$Fe sediments about 2.5 Myrs ago, which were then swiftly  transported to the surface of Earth \citep{2023ApJ...947...58E}. Since the older Tuc-Hor stellar association was also widely discussed as a possible SN candidate, we also consider it here for the sake of completeness. The observational data on cosmic-ray flux and composition  in the knee region  allows us to effectively constrain the physical parameters of our SN model and provide a realistic estimate of the radiation doses at various times since the SN explosion took place. Our results are  tellingly summarized in Figures~\ref{fig:doses1} and \ref{fig:doses2}.

The probability for a nearby SN occurrence is increased  because the Solar System recently entered the LB. Fifteen SN explosions are estimated to have occurred in order  to inflate the LB over the last 15 million years \citep{2002PhRvL..88h1101B,2023ApJ...947...58E}. We know from the  reconstruction of the LB history \citep{Zucker2022} that at least 9 SN exploded during the past 6 Myrs. This gives about one SN every $\approx  6.6 \times 10^5$ years at a distance less than 150 pc. Assuming that the filling factor of stellar clusters in the  LB is $f$,  the SN rate in the LB can then be written  as $\approx 2\times 10^{-3}f$ kpc$^{-2}$ yr$^{-1}$. This simple estimate agrees well with the  historical SN rate for $f\approx 0.1$, which gives approximately 1 event every 50 years in our Galaxy.

The results presented here for the expected cosmic-ray flux from a nearby SN differ from those described in \citet{2017ApJ...840..105M},  which we have used previously in \citet{2021ApJ...910...85G}. The reason is twofold. First, they  assume a distance of  50 pc, while we now know that Tuc-Hor was further away ($\approx 70$ pc) at the time of the SN explosion. Second, they presume $2.5 \times 10^{50}$~erg in cosmic rays and a spectral index of 2.2, with a cutoff at 1 PeV. Here we  show that to actually fit the knee region of the cosmic-ray spectrum, {    one  needs a harder spectrum to fit the data, $\alpha=1.6-1.7$, an energy in cosmic rays $N_0\approx 10^{49}$~erg  for $D_0=10^{27}$ cm$^{2}$ s$^{-1}$ ($N_0\approx 10^{50}$~erg for $D_0=10^{28}$ cm$^{2}$ s$^{-1}$, respectively), which is similar the $\gamma$-ray luminosity in SNRs, and a rigidity cutoff at 5 PV (to effectively capture the knee).} 

We also show that the spectral shape varies with time and this needs to be taken into account when calculating the corresponding doses (Figures~\ref{fig:doses1} and \ref{fig:doses2}). We find lower average doses as calculated over extended periods of time.  That is,  $\approx 10$ mGy/yr during the first  $10^4$ yrs  after a SN explosion in Tuc-Hor and  $\approx 2$ mGy/yr during  the first $10^4$ yrs after a SN explosion in Sco-Cen (UCL), {    under the assumption that $D_0=10^{27}$ cm$^{2}$ s$^{-1}$. For  $D_0=10^{28}$ cm$^{2}$ s$^{-1}$ one needs larger energy content in cosmic rays to effectively describe the observations, typically around $10^{50}$~erg. This agrees with cosmic-ray acceleration models where $\approx$10\% of the shock energy is transferred into the energy of the accelerated cosmic rays \citep[e.g.][]{2014ApJ...783...91C}. In this case, the average dose is $\approx 100$ mGy/yr during the first  $10^4$ yrs  after a SN explosion in Tuc-Hor and  $\approx 30$ mGy/yr during  the first $10^4$ yrs after a SN explosion in Sco-Cen (UCL). For completeness, in Appendix \ref{app:tur} we compare how our results vary when assuming a diffusion coefficient that is determined by either Kolmogorov's  or Kraichnan's theory of turbulence.} 

It is not clear what would the biological effects of such radiation doses be. The study of populations living in Kerala, India, where the background radiation level was observed to vary between 0.1 and 45.0 mGy/yr showed that  5.0 mGy/yr (mean dose) may be the threshold dose for double strand break induction \citep{jain2016lack}. Double strand breaks in DNA can potentially lead to mutations and jump in the diversification of species. \cite{COSTA20241247} showed that the rate of virus diversification  in the African Tanganyika lake  accelerated 2-3 Myr ago. It would be appealing to  better understand whether this can be attributed  to the increase in cosmic-radiation dose we predict to have taking place during that period. We note that the calculated dose from a SN occurring in Tuc-Hor, whose properties can  account for the $^{60}$Fe concentration peak 2.5~Myr ago as well as describe the cosmic-ray spectrum and composition in the knee region, would certainly not induce a mass extinction. On the other hand, it could lead to a diversification of species through an increase in the mutation rate.  

By way of comparison, a radiation dose for a SN occurring at 10 pc (considering the same rate as above, which gives one event approximately every 150 Myr), assuming  a diffusion coefficient of $10^{27}$~cm$^2$~s$^{-1}$ and a total energy of $10^{49}$~erg in cosmic rays, gives an average radiation dose of $\sim$500 mGy/yr, which has been calculated  averaging over the first 10 kyr. The dose limit  for occupational exposure in a nuclear facility is, for comparative purposes, 500 mSv/yr for the skin and extremities.\footnote{1 mSv/yr$\equiv$1 mGy/yr for photons and leptons.}  We present in Figure~\ref{fig:dosedist}   the average dose as a function of SN distance under the same stated assumptions. {    Even a SN at 200~pc (which is roughly our distance to Betelgeuse) would increase the cosmic ray radiation dose between $\sim$1~mGy/yr ($D_0=10^{27}$~cm$^2$~s$^{-1}$) and $\sim$30~mGy/yr ($D_0=10^{28}$~cm$^2$~s$^{-1}$), depending on the local diffusion coefficient. It is therefore important to be able to better understand the structure of our local magnetic turbulence and better constrain the value of the local diffusion coefficient to  estimate the radiation doses. It should be noted that a value of a few $D_0=10^{28}$~cm$^2$~s$^{-1}$ has usually been invoked in previous studies \citep{2018PhRvD..97f3011K, 2024MNRAS.530..684D}. We also present results for the extreme case $D_0=10^{29}$~cm$^2$~s$^{-1}$ where the  energy injected in cosmic rays has to be $\approx3\times10^{51}$~erg in order to fit the knee with a 3~Myr old explosion in Sco-Cen (note that for such a yield in cosmic rays, the PeVatron is more likely to be associated with a superluminous supernova or hypernova).}

It is therefore certain that cosmic radiation is a key environmental factor when assessing the viability and evolution of life on Earth, and the key question pertains to the threshold for radiation to be a favorable or harmful trigger when considering the evolution of species. The exact threshold can only be established with a clear understanding of the biological effects of cosmic radiation (especially muons that  dominate at ground level), which remains highly unexplored.

We finally remark that in our model  the “knee” in the cosmic-ray spectrum, which is due to a nearby SN, is an ephemeral structure\footnote{    by contrast to models where it is due to a change  in the escape mechanism of cosmic rays from the Galactic disk. }. This structure is essential for placing stringent constraints on the cosmic-ray energy content of the PeVatron source as well as on the cosmic-ray diffusion coefficient. We predict  that the anisotropy in the PeV range,  is in the direction of one of the nearby stellar associations that is responsible for hosting the nearby SN. Another forecast we make is that the direction {    and amplitude of the cosmic-ray } anisotropy should not change between 100 TeV and 100 PeV as this is the energy at which the Galactic to extragalactic transition commences. 

As such, cosmic rays {    from nearby SN} play a key role in the development of life on Earth by potentially influencing the mutation rate of early life forms and, as such, potentially assisting in the evolution of complex organisms \citep[e.g.,][]{jain2016lack} as well as  even shaping the “handedness” of biological molecules \citep{2020ApJ...895L..11G}.
\section*{Acknowledgements}
{    We thank the  referee for their constructive remarks and suggestions.} We thank Roger Blandford, Anne Kolborg, Mordecai-Mark Mac Low, Igor Moskalenko, Theo O'Neill, Mehrnoosh Tahani, Myriam Telus, Angela Twum for helpful discussions and encouragements.  C.N. is grateful to Yulianna Ortega, Xingci Situ, Felix Perez for their support and encouragement through the  years. We thank Anatoli Fedynitch for his important contributions to our previous work. We acknowledge grants by the Sloan Foundation (G-2023-19591, NG), the Simons Foundation (MP-SCMPS-00001470, NG) and the Heising-Simons Foundation. This work was catalyzed during the UCSC STEM Diversity Program and Lamat REU program (CN). UCSC STEM Diversity Program is supported through UNIVERSITY OF CALIFORNIA {\textcopyright} Regents of the University of California- UC LEADS. The Lamat REU program  is supported by NSF grant 2150255. 
\bibliography{refs}

\begin{thebibliography}{}
\expandafter\ifx\csname natexlab\endcsname\relax\def\natexlab#1{#1}\fi
\providecommand{\url}[1]{\href{#1}{#1}}
\providecommand{\dodoi}[1]{doi:~\href{http://doi.org/#1}{\nolinkurl{#1}}}
\providecommand{\doeprint}[1]{\href{http://ascl.net/#1}{\nolinkurl{http://ascl.net/#1}}}
\providecommand{\doarXiv}[1]{\href{https://arxiv.org/abs/#1}{\nolinkurl{https://arxiv.org/abs/#1}}}

\bibitem[{{Abeysekara} {et~al.}(2017){Abeysekara}, {Albert}, {Alfaro},
  {Alvarez}, {{\'A}lvarez}, {Arceo}, {Arteaga-Vel{\'a}zquez}, {Avila Rojas},
  {Ayala Solares}, {Barber}, {Bautista-Elivar}, {Becerril}, {Belmont-Moreno},
  {BenZvi}, {Berley}, {Bernal}, {Braun}, {Brisbois}, {Caballero-Mora},
  {Capistr{\'a}n}, {Carrami{\~n}ana}, {Casanova}, {Castillo}, {Cotti},
  {Cotzomi}, {Couti{\~n}o de Le{\'o}n}, {De Le{\'o}n}, {De la Fuente},
  {Dingus}, {DuVernois}, {D{\'\i}az-V{\'e}lez}, {Ellsworth}, {Engel},
  {Enr{\'\i}quez-Rivera}, {Fiorino}, {Fraija}, {Garc{\'\i}a-Gonz{\'a}lez},
  {Garfias}, {Gerhardt}, {Gonz{\'a}lez Mu{\~n}oz}, {Gonz{\'a}lez}, {Goodman},
  {Hampel-Arias}, {Harding}, {Hern{\'a}ndez}, {Hern{\'a}ndez-Almada}, {Hinton},
  {Hona}, {Hui}, {H{\"u}ntemeyer}, {Iriarte}, {Jardin-Blicq}, {Joshi},
  {Kaufmann}, {Kieda}, {Lara}, {Lauer}, {Lee}, {Lennarz}, {Vargas},
  {Linnemann}, {Longinotti}, {Luis Raya}, {Luna-Garc{\'\i}a}, {L{\'o}pez-Coto},
  {Malone}, {Marinelli}, {Martinez}, {Martinez-Castellanos},
  {Mart{\'\i}nez-Castro}, {Mart{\'\i}nez-Huerta}, {Matthews},
  {Miranda-Romagnoli}, {Moreno}, {Mostaf{\'a}}, {Nellen}, {Newbold}, {Nisa},
  {Noriega-Papaqui}, {Pelayo}, {Pretz}, {P{\'e}rez-P{\'e}rez}, {Ren}, {Rho},
  {Rivi{\`e}re}, {Rosa-Gonz{\'a}lez}, {Rosenberg}, {Ruiz-Velasco}, {Salazar},
  {Salesa Greus}, {Sandoval}, {Schneider}, {Schoorlemmer}, {Sinnis}, {Smith},
  {Springer}, {Surajbali}, {Taboada}, {Tibolla}, {Tollefson}, {Torres},
  {Ukwatta}, {Vianello}, {Weisgarber}, {Westerhoff}, {Wisher}, {Wood},
  {Yapici}, {Yodh}, {Younk}, {Zepeda}, {Zhou}, {Guo}, {Hahn}, {Li}, \&
  {Zhang}}]{2017Sci...358..911A}
{Abeysekara}, A.~U., {Albert}, A., {Alfaro}, R., {et~al.} 2017, Science, 358,
  911, \dodoi{10.1126/science.aan4880}

\bibitem[{{Ackermann} {et~al.}(2011){Ackermann}, {Ajello}, {Allafort},
  {Baldini}, {Ballet}, {Barbiellini}, {Bastieri}, {Belfiore}, {Bellazzini},
  {Berenji}, {Blandford}, {Bloom}, {Bonamente}, {Borgland}, {Bottacini},
  {Brigida}, {Bruel}, {Buehler}, {Buson}, {Caliandro}, {Cameron}, {Caraveo},
  {Casandjian}, {Cecchi}, {Chekhtman}, {Cheung}, {Chiang}, {Ciprini}, {Claus},
  {Cohen-Tanugi}, {de Angelis}, {de Palma}, {Dermer}, {do Couto e Silva},
  {Drell}, {Dumora}, {Favuzzi}, {Fegan}, {Focke}, {Fortin}, {Fukazawa},
  {Fusco}, {Gargano}, {Germani}, {Giglietto}, {Giordano}, {Giroletti},
  {Glanzman}, {Godfrey}, {Grenier}, {Guillemot}, {Guiriec}, {Hadasch},
  {Hanabata}, {Harding}, {Hayashida}, {Hayashi}, {Hays}, {J{\'o}hannesson},
  {Johnson}, {Kamae}, {Katagiri}, {Kataoka}, {Kerr}, {Kn{\"o}dlseder}, {Kuss},
  {Lande}, {Latronico}, {Lee}, {Longo}, {Loparco}, {Lott}, {Lovellette},
  {Lubrano}, {Martin}, {Mazziotta}, {McEnery}, {Mehault}, {Michelson},
  {Mitthumsiri}, {Mizuno}, {Monte}, {Monzani}, {Morselli}, {Moskalenko},
  {Murgia}, {Naumann-Godo}, {Nolan}, {Norris}, {Nuss}, {Ohsugi}, {Okumura},
  {Orlando}, {Ormes}, {Ozaki}, {Paneque}, {Parent}, {Pesce-Rollins},
  {Pierbattista}, {Piron}, {Pohl}, {Prokhorov}, {Rain{\`o}}, {Rando},
  {Razzano}, {Reposeur}, {Ritz}, {Parkinson}, {Sgr{\`o}}, {Siskind}, {Smith},
  {Spinelli}, {Strong}, {Takahashi}, {Tanaka}, {Thayer}, {Thayer}, {Thompson},
  {Tibaldo}, {Torres}, {Tosti}, {Tramacere}, {Troja}, {Uchiyama},
  {Vandenbroucke}, {Vasileiou}, {Vianello}, {Vitale}, {Waite}, {Wang}, {Winer},
  {Wood}, {Yang}, {Zimmer}, \& {Bontemps}}]{2011Sci...334.1103A}
{Ackermann}, M., {Ajello}, M., {Allafort}, A., {et~al.} 2011, Science, 334,
  1103, \dodoi{10.1126/science.1210311}

\bibitem[{{Aglietta} {et~al.}(2004){Aglietta}, {Alessandro}, {Antonioli},
  {Arneodo}, {Bergamasco}, {Bertaina}, {Castagnoli}, {Castellina}, {Chiavassa},
  {Cini}, {D'Ettorre Piazzoli}, {di Sciascio}, {Fulgione}, {Galeotti}, {Ghia},
  {Iacovacci}, {Mannocchi}, {Morello}, {Navarra}, {Saavedra}, {Stamerra},
  {Trinchero}, {Valchierotti}, {Vallania}, {Vernetto}, {Vigorito}, {Ambrosio},
  {Antolini}, {Baldini}, {Barbarino}, {Barish}, {Battistoni}, {Becherini},
  {Bellotti}, {Bemporad}, {Bernardini}, {Bilokon}, {Bower}, {Brigida},
  {Bussino}, {Cafagna}, {Calicchio}, {Campana}, {Carboni}, {Caruso},
  {Cecchini}, {Cei}, {Chiarella}, {Choudhary}, {Coutu}, {Cozzi}, {de Cataldo},
  {Dekhissi}, {de Marzo}, {de Mitri}, {Derkaoui}, {de Vincenzi}, {di Credico},
  {Erriquez}, {Favuzzi}, {Forti}, {Fusco}, {Giacomelli}, {Giannini},
  {Giglietto}, {Giorgini}, {Grassi}, {Grillo}, {Guarino}, {Gustavino}, {Habig},
  {Hanson}, {Heinz}, {Iarocci}, {Katsavounidis}, {Katsavounidis}, {Kearns},
  {Kim}, {Kyriazopoulou}, {Lamanna}, {Lane}, {Levin}, {Lipari}, {Longley},
  {Longo}, {Loparco}, {Maaroufi}, {Mancarella}, {Mandrioli}, {Margiotta},
  {Marini}, {Martello}, {Marzari-Chiesa}, {Mazziotta}, {Michael}, {Monacelli},
  {Montaruli}, {Monteno}, {Mufson}, {Musser}, {Nicol{\`o}}, {Nolty}, {Orth},
  {Osteria}, {Palamara}, {Patera}, {Patrizii}, {Pazzi}, {Peck}, {Perrone},
  {Petrera}, {Popa}, {Rain{\`o}}, {Reynoldson}, {Ronga}, {Satriano},
  {Scapparone}, {Scholberg}, {Sciubba}, {Serra}, {Sioli}, {Sirri}, {Sitta},
  {Spinelli}, {Spinetti}, {Spurio}, {Steinberg}, {Stone}, {Sulak}, {Surdo},
  {Tarl{\`e}}, {Togo}, {Vakili}, {Walter}, \& {Webb}}]{2004APh....20..641A}
{Aglietta}, M., {Alessandro}, B., {Antonioli}, P., {et~al.} 2004, Astroparticle
  Physics, 20, 641, \dodoi{10.1016/j.astropartphys.2003.10.004}

\bibitem[{{Aharonian} \& {Atoyan}(1996)}]{1996A&A...309..917A}
{Aharonian}, F.~A., \& {Atoyan}, A.~M. 1996, \aap, 309, 917

\bibitem[{{Alemanno} {et~al.}(2024){Alemanno}, {Altomare}, {An}, {Azzarello},
  {Barbato}, {Bernardini}, {Bi}, {Cagnoli}, {Cai}, {Casilli}, {Catanzani},
  {Chang}, {Chen}, {Chen}, {Chen}, {Coppin}, {Cui}, {Cui}, {Cui}, {Dai}, {de
  Benedittis}, {de Mitri}, {de Palma}, {Deliyergiyev}, {di Giovanni}, {di
  Santo}, {Ding}, {Dong}, {Dong}, {Donvito}, {Droz}, {Duan}, {Duan}, {Fan},
  {Fan}, {Fang}, {Fang}, {Feng}, {Feng}, {Fernandez Alonso}, {Frieden},
  {Fusco}, {Gao}, {Gargano}, {Gong}, {Gong}, {Guo}, {Guo}, {Han}, {Hu},
  {Huang}, {Huang}, {Huang}, {Ionica}, {Jiang}, {Jiang}, {Jiang}, {Kong},
  {Kotenko}, {Kyratzis}, {Lei}, {Li}, {Li}, {Li}, {Li}, {Liang}, {Liu}, {Liu},
  {Liu}, {Liu}, {Liu}, {Loparco}, {Luo}, {Ma}, {Ma}, {Ma}, {Ma}, {Marsella},
  {Mazziotta}, {Mo}, {Salinas}, {Niu}, {Pan}, {Parenti}, {Peng}, {Peng},
  {Perrina}, {Putti-Garcia}, {Qiao}, {Rao}, {Ruina}, {Shangguan}, {Shen},
  {Shen}, {Shen}, {Silveri}, {Song}, {Stolpovskiy}, {Su}, {Su}, {Sun}, {Sun},
  {Surdo}, {Teng}, {Tykhonov}, {Wang}, {Wang}, {Wang}, {Wang}, {Wang}, {Wang},
  {Wang}, {Wang}, {Wang}, {Wei}, {Wei}, {Wei}, {Wu}, {Wu}, {Wu}, {Wu}, {Wu},
  {Xia}, {Xu}, {Xu}, {Xu}, {Xu}, {Xu}, {Xu}, {Xue}, {Yang}, {Yang}, {Yang},
  {Yao}, {Yu}, {Yuan}, {Yuan}, {Yue}, {Zang}, {Zhang}, {Zhang}, {Zhang},
  {Zhang}, {Zhang}, {Zhang}, {Zhang}, {Zhang}, {Zhang}, {Zhang}, {Zhao},
  {Zhao}, {Zhao}, {Zhou}, {Zhu}, \& {Dampe
  Collaboration}}]{2024PhRvD.109l1101A}
{Alemanno}, F., {Altomare}, C., {An}, Q., {et~al.} 2024, \prd, 109, L121101,
  \dodoi{10.1103/PhysRevD.109.L121101}

\bibitem[{{Allen}(1999)}]{1999ICRC....3..480A}
{Allen}, G. 1999, in International Cosmic Ray Conference, Vol.~3, 26th
  International Cosmic Ray Conference (ICRC26), Volume 3, 480,
  \dodoi{10.48550/arXiv.astro-ph/9908209}

\bibitem[{{Amenomori} {et~al.}(2008){Amenomori}, {Bi}, {Chen}, {Cui},
  {Danzengluobu}, {Ding}, {Ding}, {Fan}, {Feng}, {Feng}, {Feng}, {Gao}, {Geng},
  {Guo}, {He}, {He}, {Hibino}, {Hotta}, {Hu}, {Hu}, {Huang}, {Huang}, {Jia},
  {Kajino}, {Kasahara}, {Katayose}, {Kato}, {Kawata}, {Labaciren}, {Le}, {Li},
  {Li}, {Lou}, {Lu}, {Lu}, {Meng}, {Mizutani}, {Mu}, {Munakata}, {Nagai},
  {Nanjo}, {Nishizawa}, {Ohnishi}, {Ohta}, {Onuma}, {Ouchi}, {Ozawa}, {Ren},
  {Saito}, {Saito}, {Sakata}, {Sako}, {Shibata}, {Shiomi}, {Shirai},
  {Sugimoto}, {Takita}, {Tan}, {Tateyama}, {Torii}, {Tsuchiya}, {Udo}, {Wang},
  {Wang}, {Wang}, {Wang}, {Wang}, {Wu}, {Xue}, {Yamamoto}, {Yan}, {Yang},
  {Yasue}, {Ye}, {Yu}, {Yuan}, {Yuda}, {Zhang}, {Zhang}, {Zhang}, {Zhang},
  {Zhang}, {Zhang}, {Zhaxisangzhu}, {Zhou}, \& {Tibet AS{\ensuremath{\gamma}}
  Collaboration}}]{2008ApJ...678.1165A}
{Amenomori}, M., {Bi}, X.~J., {Chen}, D., {et~al.} 2008, \apj, 678, 1165,
  \dodoi{10.1086/529514}

\bibitem[{{Antoni} {et~al.}(2005){Antoni}, {Apel}, {Badea}, {Bekk}, {Bercuci},
  {Bl{\"u}mer}, {Bozdog}, {Brancus}, {Chilingarian}, {Daumiller}, {Doll},
  {Engel}, {Engler}, {Fe{\ss}ler}, {Gils}, {Glasstetter}, {Haungs}, {Heck},
  {H{\"o}randel}, {Kampert}, {Klages}, {Maier}, {Mathes}, {Mayer}, {Milke},
  {M{\"u}ller}, {Obenland}, {Oehlschl{\"a}ger}, {Ostapchenko}, {Petcu},
  {Rebel}, {Risse}, {Risse}, {Roth}, {Schatz}, {Schieler}, {Scholz}, {Thouw},
  {Ulrich}, {van Buren}, {Vardanyan}, {Weindl}, {Wochele}, \&
  {Zabierowski}}]{2005APh....24....1A}
{Antoni}, T., {Apel}, W.~D., {Badea}, A.~F., {et~al.} 2005, Astroparticle
  Physics, 24, 1, \dodoi{10.1016/j.astropartphys.2005.04.001}

\bibitem[{{Apel} {et~al.}(2011){Apel}, {Arteaga-Vel{\'a}zquez}, {Bekk},
  {Bertaina}, {Bl{\"u}mer}, {Bozdog}, {Brancus}, {Buchholz}, {Cantoni},
  {Chiavassa}, {Cossavella}, {Daumiller}, {de Souza}, {di Pierro}, {Doll},
  {Engel}, {Engler}, {Finger}, {Fuhrmann}, {Ghia}, {Gils}, {Glasstetter},
  {Grupen}, {Haungs}, {Heck}, {H{\"o}randel}, {Huber}, {Huege}, {Isar},
  {Kampert}, {Kang}, {Klages}, {Link}, {{\L}uczak}, {Ludwig}, {Mathes},
  {Mayer}, {Melissas}, {Milke}, {Mitrica}, {Morello}, {Navarra},
  {Oehlschl{\"a}ger}, {Ostapchenko}, {Over}, {Palmieri}, {Petcu}, {Pierog},
  {Rebel}, {Roth}, {Schieler}, {Schr{\"o}der}, {Sima}, {Toma}, {Trinchero},
  {Ulrich}, {Weindl}, {Wochele}, {Wommer}, \&
  {Zabierowski}}]{2011PhRvL.107q1104A}
{Apel}, W.~D., {Arteaga-Vel{\'a}zquez}, J.~C., {Bekk}, K., {et~al.} 2011, \prl,
  107, 171104, \dodoi{10.1103/PhysRevLett.107.171104}

\bibitem[{{Bartoli} {et~al.}(2015){Bartoli}, {Bernardini}, {Bi}, {Cao},
  {Catalanotti}, {Chen}, {Chen}, {Cui}, {Dai}, {D'Amone}, {Danzengluobu}, {De
  Mitri}, {D'Ettorre Piazzoli}, {Di Girolamo}, {Di Sciascio}, {Feng}, {Feng},
  {Feng}, {Guo}, {Guo}, {He}, {Hu}, {Hu}, {Iacovacci}, {Iuppa}, {Jia},
  {Labaciren}, {Li}, {Liu}, {Liu}, {Liu}, {Lu}, {Ma}, {Ma}, {Mancarella},
  {Mari}, {Marsella}, {Mastroianni}, {Montini}, {Ning}, {Perrone}, {Pistilli},
  {Salvini}, {Santonico}, {Shen}, {Sheng}, {Shi}, {Surdo}, {Tan}, {Vallania},
  {Vernetto}, {Vigorito}, {Wang}, {Wu}, {Wu}, {Xue}, {Yang}, {Yang}, {Yao},
  {Yuan}, {Zha}, {Zhang}, {Zhang}, {Zhang}, {Zhang}, {Zhao}, {Zhaxiciren},
  {Zhaxisangzhu}, {Zhou}, {Zhu}, {Zhu}, {Bai}, {Chen}, {Feng}, {Gao}, {Gu},
  {Hou}, {Liu}, {Liu}, {Wang}, {Xiao}, {Zhang}, {Zhang}, {Zhou}, {Zuo}, \&
  {ARGO-YBJ Collaboration}}]{2015PhRvD..92i2005B}
{Bartoli}, B., {Bernardini}, P., {Bi}, X.~J., {et~al.} 2015, \prd, 92, 092005,
  \dodoi{10.1103/PhysRevD.92.092005}

\bibitem[{{Ben{\'\i}tez} {et~al.}(2002){Ben{\'\i}tez},
  {Ma{\'\i}z-Apell{\'a}niz}, \& {Canelles}}]{2002PhRvL..88h1101B}
{Ben{\'\i}tez}, N., {Ma{\'\i}z-Apell{\'a}niz}, J., \& {Canelles}, M. 2002,
  \prl, 88, 081101, \dodoi{10.1103/PhysRevLett.88.081101}

\bibitem[{Berger {et~al.}(2017)Berger, Coursey, Zucker, \&
  Chang}]{ionization_tables}
Berger, M.~J., Coursey, J., Zucker, M., \& Chang, J. 2017, NIST Standard
  Reference Database 124 (Version 2.0.1), \dodoi{10.18434/T4NC7P}

\bibitem[{{Caprioli} \& {Spitkovsky}(2014)}]{2014ApJ...783...91C}
{Caprioli}, D., \& {Spitkovsky}, A. 2014, \apj, 783, 91,
  \dodoi{10.1088/0004-637X/783/2/91}

\bibitem[{Costa {et~al.}(2024)Costa, Ronco, Mifsud, Harvey, Salzburger, \&
  Holmes}]{COSTA20241247}
Costa, V.~A., Ronco, F., Mifsud, J.~C., {et~al.} 2024, Current Biology, 34,
  1247, \dodoi{https://doi.org/10.1016/j.cub.2024.02.008}

\bibitem[{{Cristofari}(2021)}]{2021Univ....7..324C}
{Cristofari}, P. 2021, Universe, 7, 324, \dodoi{10.3390/universe7090324}

\bibitem[{{Dampe Collaboration}(2022)}]{2022SciBu..67.2162D}
{Dampe Collaboration}. 2022, Science Bulletin, 67, 2162,
  \dodoi{10.1016/j.scib.2022.10.002}

\bibitem[{{de S{\'e}r{\'e}ville} {et~al.}(2024){de S{\'e}r{\'e}ville},
  {Tatischeff}, {Cristofari}, {Gabici}, \& {Diehl}}]{2024MNRAS.530..684D}
{de S{\'e}r{\'e}ville}, N., {Tatischeff}, V., {Cristofari}, P., {Gabici}, S.,
  \& {Diehl}, R. 2024, \mnras, 530, 684, \dodoi{10.1093/mnras/stae336}

\bibitem[{{Dembinski} {et~al.}(2017){Dembinski}, {Engel}, {Fedynitch},
  {Gaisser}, {Riehn}, \& {Stanev}}]{2017ICRC...35..533D}
{Dembinski}, H., {Engel}, R., {Fedynitch}, A., {et~al.} 2017, in International
  Cosmic Ray Conference, Vol. 301, 35th International Cosmic Ray Conference
  (ICRC2017), 533, \dodoi{10.22323/1.301.0533}

\bibitem[{{Egger} \& {Aschenbach}(1995)}]{1995A&A...294L..25E}
{Egger}, R.~J., \& {Aschenbach}, B. 1995, \aap, 294, L25,
  \dodoi{10.48550/arXiv.astro-ph/9412086}

\bibitem[{{Ellis} \& {Schramm}(1995)}]{1995PNAS...92..235E}
{Ellis}, J., \& {Schramm}, D.~N. 1995, Proceedings of the National Academy of
  Science, 92, 235, \dodoi{10.1073/pnas.92.1.235}

\bibitem[{{Ertel} {et~al.}(2023){Ertel}, {Fry}, {Fields}, \&
  {Ellis}}]{2023ApJ...947...58E}
{Ertel}, A.~F., {Fry}, B.~J., {Fields}, B.~D., \& {Ellis}, J. 2023, \apj, 947,
  58, \dodoi{10.3847/1538-4357/acb699}

\bibitem[{Fang {et~al.}(2020)Fang, Bi, \& Yin}]{Fang_2020}
Fang, K., Bi, X.-J., \& Yin, P.-F. 2020, The Astrophysical Journal, 903, 69,
  \dodoi{10.3847/1538-4357/abb8d7}

\bibitem[{{Fedynitch} {et~al.}(2015){Fedynitch}, {Engel}, {Gaisser}, {Riehn},
  \& {Stanev}}]{2015EPJWC..9908001F}
{Fedynitch}, A., {Engel}, R., {Gaisser}, T.~K., {Riehn}, F., \& {Stanev}, T.
  2015, in European Physical Journal Web of Conferences, Vol.~99, European
  Physical Journal Web of Conferences, 08001,
  \dodoi{10.1051/epjconf/20159908001}

\bibitem[{Frisch {et~al.}(2011)Frisch, Redfield, \& Slavin}]{Frisch2011}
Frisch, P.~C., Redfield, S., \& Slavin, J.~D. 2011, Annual Review of Astronomy
  and Astrophysics, 49, 237, \dodoi{10.1146/annurev-astro-081710-102613}

\bibitem[{{Fry} {et~al.}(2015){Fry}, {Fields}, \&
  {Ellis}}]{2015ApJ...800...71F}
{Fry}, B.~J., {Fields}, B.~D., \& {Ellis}, J.~R. 2015, \apj, 800, 71,
  \dodoi{10.1088/0004-637X/800/1/71}

\bibitem[{{Fuchs} {et~al.}(2006){Fuchs}, {Breitschwerdt}, {de Avillez},
  {Dettbarn}, \& {Flynn}}]{2006MNRAS.373..993F}
{Fuchs}, B., {Breitschwerdt}, D., {de Avillez}, M.~A., {Dettbarn}, C., \&
  {Flynn}, C. 2006, \mnras, 373, 993, \dodoi{10.1111/j.1365-2966.2006.11044.x}

\bibitem[{{Fujii}(2024)}]{2024arXiv240108952F}
{Fujii}, T. 2024, arXiv e-prints, arXiv:2401.08952,
  \dodoi{10.48550/arXiv.2401.08952}

\bibitem[{{Gallegos-Garcia} {et~al.}(2020){Gallegos-Garcia}, {Burkhart},
  {Rosen}, {Naiman}, \& {Ramirez-Ruiz}}]{2020ApJ...899L..30G}
{Gallegos-Garcia}, M., {Burkhart}, B., {Rosen}, A.~L., {Naiman}, J.~P., \&
  {Ramirez-Ruiz}, E. 2020, \apjl, 899, L30, \dodoi{10.3847/2041-8213/ababae}

\bibitem[{{Galli} {et~al.}(2023){Galli}, {Miret-Roig}, {Bouy}, {Olivares}, \&
  {Barrado}}]{2023MNRAS.520.6245G}
{Galli}, P. A.~B., {Miret-Roig}, N., {Bouy}, H., {Olivares}, J., \& {Barrado},
  D. 2023, \mnras, 520, 6245, \dodoi{10.1093/mnras/stad520}

\bibitem[{{Globus} {et~al.}(2015){Globus}, {Allard}, \&
  {Parizot}}]{2015PhRvD..92b1302G}
{Globus}, N., {Allard}, D., \& {Parizot}, E. 2015, \prd, 92, 021302,
  \dodoi{10.1103/PhysRevD.92.021302}

\bibitem[{{Globus} \& {Blandford}(2020)}]{2020ApJ...895L..11G}
{Globus}, N., \& {Blandford}, R.~D. 2020, \apjl, 895, L11,
  \dodoi{10.3847/2041-8213/ab8dc6}

\bibitem[{{Globus} {et~al.}(2021){Globus}, {Fedynitch}, \&
  {Blandford}}]{2021ApJ...910...85G}
{Globus}, N., {Fedynitch}, A., \& {Blandford}, R.~D. 2021, \apj, 910, 85,
  \dodoi{10.3847/1538-4357/abe461}

\bibitem[{{Herbst} {et~al.}(2012){Herbst}, {Heber}, {Kopp}, {Sternal}, \&
  {Steinhilber}}]{2012ApJ...761...17H}
{Herbst}, K., {Heber}, B., {Kopp}, A., {Sternal}, O., \& {Steinhilber}, F.
  2012, \apj, 761, 17, \dodoi{10.1088/0004-637X/761/1/17}

\bibitem[{{Hoogerwerf} {et~al.}(2000){Hoogerwerf}, {de Bruijne}, \& {de
  Zeeuw}}]{2000ApJ...544L.133H}
{Hoogerwerf}, R., {de Bruijne}, J.~H.~J., \& {de Zeeuw}, P.~T. 2000, \apjl,
  544, L133, \dodoi{10.1086/317315}

\bibitem[{{H{\"o}randel}(2003)}]{2003APh....19..193H}
{H{\"o}randel}, J.~R. 2003, Astroparticle Physics, 19, 193,
  \dodoi{10.1016/S0927-6505(02)00198-6}

\bibitem[{Hyde \& Pecaut(2018)}]{HydePecaut}
Hyde, M., \& Pecaut, M.~J. 2018, Astronomische Nachrichten, 339, 78,
  \dodoi{https://doi.org/10.1002/asna.201713375}

\bibitem[{Jain {et~al.}(2016)Jain, Kumar, Koya, Jaikrishan, \&
  Das}]{jain2016lack}
Jain, V., Kumar, P.~V., Koya, P., Jaikrishan, G., \& Das, B. 2016, Mutation
  Research/Fundamental and Molecular Mechanisms of Mutagenesis, 788, 50

\bibitem[{{Kachelrie{\ss}} {et~al.}(2018){Kachelrie{\ss}}, {Neronov}, \&
  {Semikoz}}]{2018PhRvD..97f3011K}
{Kachelrie{\ss}}, M., {Neronov}, A., \& {Semikoz}, D.~V. 2018, \prd, 97,
  063011, \dodoi{10.1103/PhysRevD.97.063011}

\bibitem[{{Kang} {et~al.}(2023){Kang}, {Arteaga-Vel{\'a}zquez}, {Bertaina},
  {Chiavassa}, {Daumiller}, {de Souza}, {Engel}, {Gherghel-Lascu}, {Grupen},
  {Haungs}, {H{\"o}randel}, {Huege}, {Kampert}, {Link}, {Mathes},
  {Ostapchenko}, {Pierog}, {Rivera-Rangel}, {Roth}, {Schieler}, {Schr{\"o}der},
  {Sima}, {Weindl}, {Wochele}, \& {Zabierowski}}]{2023arXiv231205054K}
{Kang}, D., {Arteaga-Vel{\'a}zquez}, J.~C., {Bertaina}, M., {et~al.} 2023,
  arXiv e-prints, arXiv:2312.05054, \dodoi{10.48550/arXiv.2312.05054}

\bibitem[{Karam \& Leslie(1999)}]{karam1999calculations}
Karam, P.~A., \& Leslie, S.~A. 1999, Health physics, 77, 662

\bibitem[{{Kolborg} {et~al.}(2022){Kolborg}, {Martizzi}, {Ramirez-Ruiz},
  {Pfister}, {Sakari}, {Wechsler}, \& {Soares-Furtado}}]{2022ApJ...936L..26K}
{Kolborg}, A.~N., {Martizzi}, D., {Ramirez-Ruiz}, E., {et~al.} 2022, \apjl,
  936, L26, \dodoi{10.3847/2041-8213/ac8c98}

\bibitem[{{Kolborg} {et~al.}(2023){Kolborg}, {Ramirez-Ruiz}, {Martizzi},
  {Macias}, \& {Soares-Furtado}}]{2023ApJ...949..100K}
{Kolborg}, A.~N., {Ramirez-Ruiz}, E., {Martizzi}, D., {Macias}, P., \&
  {Soares-Furtado}, M. 2023, \apj, 949, 100, \dodoi{10.3847/1538-4357/acca80}

\bibitem[{{Kuznetsov} {et~al.}(2024){Kuznetsov}, {Petrov}, {Plokhikh}, \&
  {Sotnikov}}]{2024JCAP...05..125K}
{Kuznetsov}, M.~Y., {Petrov}, N.~A., {Plokhikh}, I.~A., \& {Sotnikov}, V.~V.
  2024, \jcap, 2024, 125, \dodoi{10.1088/1475-7516/2024/05/125}

\bibitem[{{Lhaaso Collaboration}(2024)}]{2024SciBu..69..449L}
{Lhaaso Collaboration}. 2024, Science Bulletin, 69, 449,
  \dodoi{10.1016/j.scib.2023.12.040}

\bibitem[{{Lingenfelter}(2018)}]{2018AdSpR..62.2750L}
{Lingenfelter}, R.~E. 2018, Advances in Space Research, 62, 2750,
  \dodoi{10.1016/j.asr.2017.04.006}

\bibitem[{{Ma{\'\i}z-Apell{\'a}niz}(2001)}]{2001ApJ...560L..83M}
{Ma{\'\i}z-Apell{\'a}niz}, J. 2001, \apjl, 560, L83, \dodoi{10.1086/324016}

\bibitem[{{Malkov} \& {Moskalenko}(2022)}]{2022ApJ...933...78M}
{Malkov}, M.~A., \& {Moskalenko}, I.~V. 2022, \apj, 933, 78,
  \dodoi{10.3847/1538-4357/ac7049}

\bibitem[{{Mamajek}(2016)}]{2016IAUS..314...21M}
{Mamajek}, E.~E. 2016, in Young Stars \& Planets Near the Sun, ed. J.~H.
  {Kastner}, B.~{Stelzer}, \& S.~A. {Metchev}, Vol. 314, 21--26,
  \dodoi{10.1017/S1743921315006250}

\bibitem[{{Meighen-Berger} \& {Li}(2019)}]{2019ICRC...36..961M}
{Meighen-Berger}, S., \& {Li}, M. 2019, in International Cosmic Ray Conference,
  Vol.~36, 36th International Cosmic Ray Conference (ICRC2019), 961.
\newblock \doarXiv{1910.05984}

\bibitem[{{Melott} {et~al.}(2017){Melott}, {Thomas}, {Kachelrie{\ss}},
  {Semikoz}, \& {Overholt}}]{2017ApJ...840..105M}
{Melott}, A.~L., {Thomas}, B.~C., {Kachelrie{\ss}}, M., {Semikoz}, D.~V., \&
  {Overholt}, A.~C. 2017, \apj, 840, 105, \dodoi{10.3847/1538-4357/aa6c57}

\bibitem[{{Morales-Soto} \&
  {Arteaga-Vel{\'a}zquez}(2022)}]{2022arXiv220814245M}
{Morales-Soto}, J.~A., \& {Arteaga-Vel{\'a}zquez}, J.~C. 2022, arXiv e-prints,
  arXiv:2208.14245, \dodoi{10.48550/arXiv.2208.14245}

\bibitem[{{Nava} {et~al.}(2019){Nava}, {Recchia}, {Gabici}, {Marcowith},
  {Brahimi}, \& {Ptuskin}}]{2019MNRAS.484.2684N}
{Nava}, L., {Recchia}, S., {Gabici}, S., {et~al.} 2019, \mnras, 484, 2684,
  \dodoi{10.1093/mnras/stz137}

\bibitem[{{Neuh{\"a}user} {et~al.}(2020){Neuh{\"a}user}, {Gie{\ss}ler}, \&
  {Hambaryan}}]{2020MNRAS.498..899N}
{Neuh{\"a}user}, R., {Gie{\ss}ler}, F., \& {Hambaryan}, V.~V. 2020, \mnras,
  498, 899, \dodoi{10.1093/mnras/stz2629}

\bibitem[{{Nimmo} {et~al.}(2020){Nimmo}, {Primack}, {Faber}, {Ramirez-Ruiz}, \&
  {Safarzadeh}}]{2020ApJ...903L..37N}
{Nimmo}, F., {Primack}, J., {Faber}, S.~M., {Ramirez-Ruiz}, E., \&
  {Safarzadeh}, M. 2020, \apjl, 903, L37, \dodoi{10.3847/2041-8213/abc251}

\bibitem[{{O'Neill} {et~al.}(2024){O'Neill}, {Zucker}, {Goodman}, \&
  {Edenhofer}}]{2024ApJ...973..136O}
{O'Neill}, T.~J., {Zucker}, C., {Goodman}, A.~A., \& {Edenhofer}, G. 2024,
  \apj, 973, 136, \dodoi{10.3847/1538-4357/ad61de}

\bibitem[{{Particle Data Group}(2020)}]{2020PTEP.2020h3C01P}
{Particle Data Group}. 2020, Progress of Theoretical and Experimental Physics,
  2020, 083C01, \dodoi{10.1093/ptep/ptaa104}

\bibitem[{{Piecka} {et~al.}(2024){Piecka}, {Hutschenreuter}, \&
  {Alves}}]{2024arXiv240713226P}
{Piecka}, M., {Hutschenreuter}, S., \& {Alves}, J. 2024, arXiv e-prints,
  arXiv:2407.13226, \dodoi{10.48550/arXiv.2407.13226}

\bibitem[{{Ratzenb{\"o}ck} {et~al.}(2023){Ratzenb{\"o}ck}, {Gro{\ss}schedl},
  {Alves}, {Miret-Roig}, {Bomze}, {Forbes}, {Goodman}, {Hacar}, {Lin},
  {Meingast}, {M{\"o}ller}, {Piecka}, {Posch}, {Rottensteiner}, {Swiggum}, \&
  {Zucker}}]{2023A&A...678A..71R}
{Ratzenb{\"o}ck}, S., {Gro{\ss}schedl}, J.~E., {Alves}, J., {et~al.} 2023,
  \aap, 678, A71, \dodoi{10.1051/0004-6361/202346901}

\bibitem[{{Robitaille} {et~al.}(2018){Robitaille}, {Scaife}, {Carretti},
  {Haverkorn}, {Crocker}, {Kesteven}, {Poppi}, \&
  {Staveley-Smith}}]{2018A&A...617A.101R}
{Robitaille}, J.~F., {Scaife}, A.~M.~M., {Carretti}, E., {et~al.} 2018, \aap,
  617, A101, \dodoi{10.1051/0004-6361/201833358}

\bibitem[{{Rosen} {et~al.}(2014){Rosen}, {Lopez}, {Krumholz}, \&
  {Ramirez-Ruiz}}]{2014MNRAS.442.2701R}
{Rosen}, A.~L., {Lopez}, L.~A., {Krumholz}, M.~R., \& {Ramirez-Ruiz}, E. 2014,
  \mnras, 442, 2701, \dodoi{10.1093/mnras/stu1037}

\bibitem[{{Schroer} {et~al.}(2022){Schroer}, {Pezzi}, {Caprioli}, {Haggerty},
  \& {Blasi}}]{2022MNRAS.512..233S}
{Schroer}, B., {Pezzi}, O., {Caprioli}, D., {Haggerty}, C.~C., \& {Blasi}, P.
  2022, \mnras, 512, 233, \dodoi{10.1093/mnras/stac466}

\bibitem[{{Schulreich} {et~al.}(2023){Schulreich}, {Feige}, \&
  {Breitschwerdt}}]{2023A&A...680A..39S}
{Schulreich}, M.~M., {Feige}, J., \& {Breitschwerdt}, D. 2023, \aap, 680, A39,
  \dodoi{10.1051/0004-6361/202347532}

\bibitem[{{Shklovskij}(1969)}]{1969supe.book.....S}
{Shklovskij}, I.~S. 1969, {Supernovae.}

\bibitem[{Standard~Atmosphere(1976)}]{us_std_atmosphere}
Standard~Atmosphere, U.~S. 1976, US Gov. Print. Off., Washington, DC

\bibitem[{{Vieu} \& {Reville}(2023)}]{Vieu2023}
{Vieu}, T., \& {Reville}, B. 2023, \mnras, 519, 136,
  \dodoi{10.1093/mnras/stac3469}

\bibitem[{Wallner {et~al.}(2021)Wallner, Froehlich, Hotchkis, Kinoshita, Paul,
  Martschini, Pavetich, Tims, Kivel, Schumann, Honda, Matsuzaki, \&
  Yamagata}]{doi:10.1126/science.aax3972}
Wallner, A., Froehlich, M.~B., Hotchkis, M. A.~C., {et~al.} 2021, Science, 372,
  742, \dodoi{10.1126/science.aax3972}

\bibitem[{{Weaver} {et~al.}(1977){Weaver}, {McCray}, {Castor}, {Shapiro}, \&
  {Moore}}]{1977ApJ...218..377W}
{Weaver}, R., {McCray}, R., {Castor}, J., {Shapiro}, P., \& {Moore}, R. 1977,
  \apj, 218, 377, \dodoi{10.1086/155692}

\bibitem[{{Zucker} {et~al.}(2022){Zucker}, {Goodman}, {Alves}, {Bialy},
  {Foley}, {Speagle}, {Gro{\^I}{\texttwosuperior}schedl}, {Finkbeiner},
  {Burkert}, {Khimey}, \& {Swiggum}}]{Zucker2022}
{Zucker}, C., {Goodman}, A.~A., {Alves}, J., {et~al.} 2022, \nat, 601, 334,
  \dodoi{10.1038/s41586-021-04286-5}

\end{thebibliography}

\appendix

\section{    Comparison between Kolmogorov and Kraichnan turbulence}
\label{app:tur}

\begin{figure*}[!h]
\begin{center}

\includegraphics[height=15.5cm]{Tuc_Hor_model27_new.png}
\includegraphics[height=15.5cm]{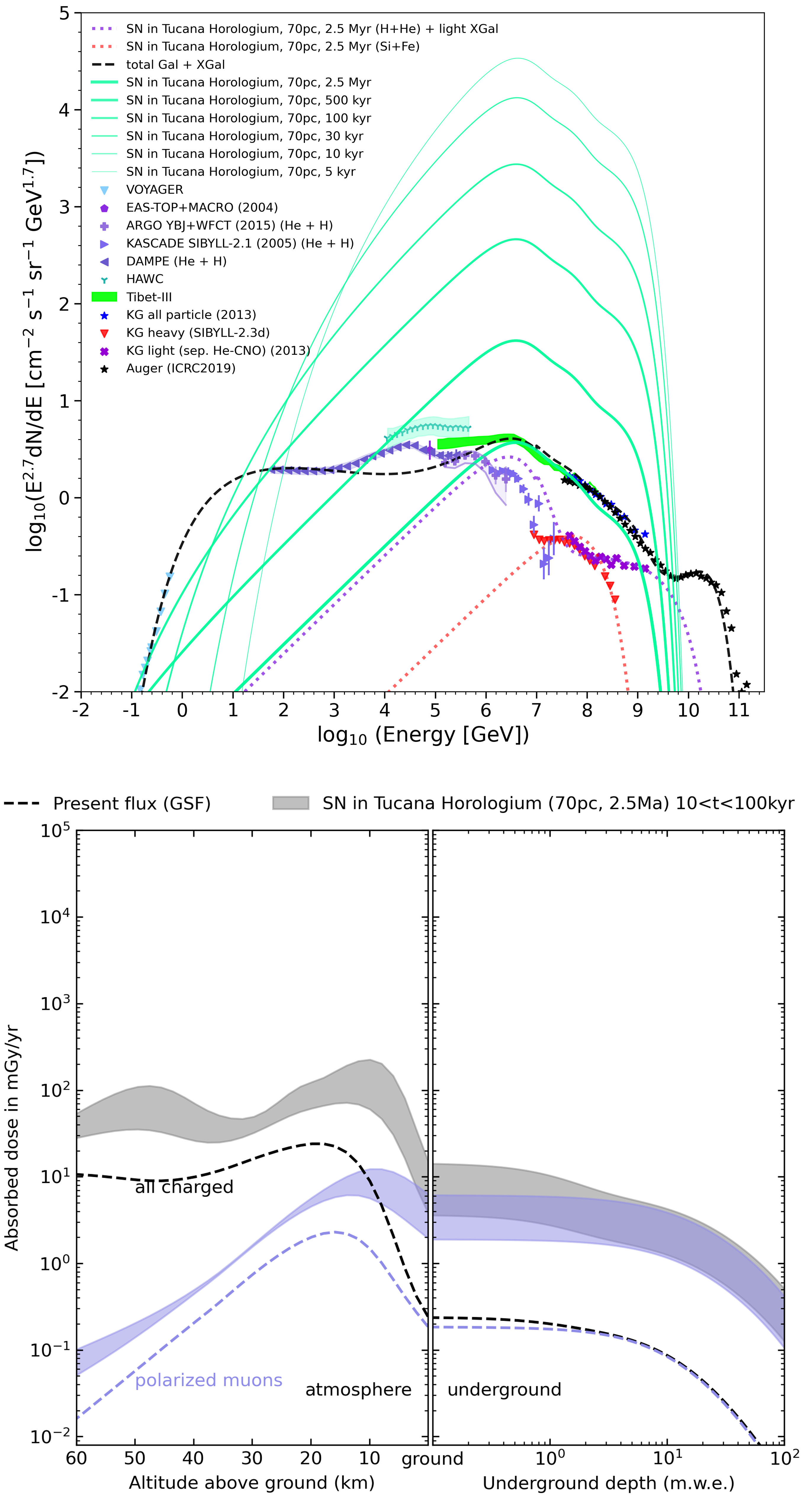}
    \caption{    The cosmic-ray spectrum ({\it upper panels}) and the corresponding radiation dose experienced at different altitudes/depths ({\it lower panels}) from a supernova explosion associated with the 2.5Myr  old $^{60}$Fe peak deposits. The parameters of the SN are: distance 70 pc, explosion time 2.5 Myr. The parameters of the source spectrum are: $N_0=10^{49}$~erg, $\alpha=1.7$, $R_{\rm cut}=5$~PV.  
    {\it Left panels}: $D_0=10^{27}$ cm$^{2}$ s$^{-1}$, $\delta=0.5$ (Kraichnan). This is the same as the left panel of Fig.~\ref{fig:doses1}.  
     Composition: $\approx$ 90\% light (H+He), $\approx$ 8.8\% CNO, $\approx1$\% Si+Fe, and $\approx 0.2$\% r-process elements. {\it Right panels}: $D_0=10^{27.85}$ cm$^{2}$ s$^{-1}$, $\delta=0.33$ (Kolmogorov). Composition: $\approx$ 92\% light (H+He), $\approx$ 7\% CNO, $\approx0.8$\% Si+Fe, and $\approx 0.2$\% r-process elements. Note that we assumed a {\it slightly} different composition to ensure a good fit of the elemental contributions.}
    \label{fig:comparison}
    \end{center}
\end{figure*}

We show in Figure~\ref{fig:comparison} how our results vary when assuming a diffusion coefficient
that is determined by either Kolmogorov's or Kraichnan's theory of turbulence.

\end{document}